\documentclass[twocolumn,showpacs,preprintnumbers,amsmath,amssymb]{revtex4}
\usepackage{graphicx}% Include figure files
\usepackage{dcolumn}% Align table columns on decimal point
\usepackage{bm}% bold math
\usepackage{longtable}
\newcommand{\eq}{eqnarray}
\newcommand{\f}{\frac}
\newcommand{\bfll}{\begin{flushleft}}
\newcommand{\efll}{\end{flushleft}}
\newcommand{\bt}{\begin{tabular}}
\newcommand{\et}{\end{tabular}}
\newcommand{\bce}{\begin{center}}
\newcommand{\ece}{\end{center}}
\newcommand{\ben}{\begin{enumerate}}
\newcommand{\een}{\end{enumerate}}
\newcommand{\be}{\begin{equation}}
\newcommand{\ee}{\end{equation}}
\newcommand{\lk}{\left(}
\newcommand{\rk}{\right)}
\newcommand{\ltk}{\left\{}
\newcommand{\rtk}{\right\}}
\newcommand{\ldk}{\left[}
\newcommand{\rdk}{\right]}

\newcommand{\nn}{\nonumber \\}

\newcommand{\sxy}{\sigma_{xy}}
\newcommand{\sxx}{\sigma_{xx}}
\newcommand{\w}{\omega}
\newcommand{\g}{\gamma}
\newcommand{\ld}{\lambda}

\newcommand{\bk}{\bm{k}}
\newcommand{\bq}{\bm{q}}
\newcommand{\br}{\bm{r}}
\newcommand{\bR}{\bm{R}}
\newcommand{\kt}{\tilde \bk}

\newcommand{\rd}{\partial}
\newcommand{\e}{\text{e}}
\newcommand{\dm}{\Delta \mu}
\begin{document}

\title{Study of Intrinsic Spin Hall Effect and Orbital Hall Effect \\
in 4$d$- and 5$d$- Transition Metals}
 
\author{
 T. {\sc Tanaka}$^1$, H. {\sc Kontani}$^1$, M. {\sc Naito}$^1$, T. {\sc Naito}$^1$, D.S. {\sc Hirashima}$^1$, K. {\sc Yamada}$^2$ and J. {\sc Inoue}$^3$
}

\address{ 
$^1$Department of Physics, Nagoya University, Furo-cho, Nagoya 464-8602, Japan. \\
$^2$Engineering, Ritsumeikan University, 1-1-1 Noji Higashi, Kusastu, 
Shiga 525-8577, Japan. \\
$^3$Department of Applied Physics, Nagoya University, Furo-cho, Nagoya 464-8602, Japan. }

\date{\today}

\begin{abstract}
We study the intrinsic spin Hall conductivity (SHC) in 
various $5d$-transition metals (Ta, W, Re, Os, Ir, Pt, and Au)
and $4d$-transition metals (Nb, Mo, Tc, Ru, Rh, Pd, and Ag)
based on the Naval Research Laboratory tight-binding model, which enables us to perform quantitatively reliable analysis.
In each metal, the obtained intrinsic SHC is independent of resistivity in the low resistive regime ($\rho < 50 \mu\Omega\text{cm}$) whereas it decreases in proportion to $\rho^{-2}$ in the high resistive regime.
In the low resistive regime, the SHC takes a large positive value in 
Pt and Pd, both of which have approximately nine $d$-electrons per ion ($n_d=9$).
On the other hand, the SHC takes a large negative value in 
Ta, Nb, W, and Mo where $n_d<5$.
%Among them, Pt shows the largest SHC in magnitude in the low resistive regime.
%We also find that the SHC is sensitive to the position of the 
%Fermi level, which can be controlled by composing alloys.
In transition metals, a conduction electron acquires the trajectory-dependent
phase factor that originates from the atomic wavefunction.
This phase factor, which is reminiscent of the Aharonov-Bohm phase, 
is the origin of the SHC in paramagnetic metals
and that of the anomalous Hall conductivity in ferromagnetic metals.
%The origin of the huge SHC in transition metals is the 
%effective Aharonov-Bohm phase associated with 
%the interorbital motion of electrons, originating from the phase factor 
%of the $d$-electron wavefunction and the atomic spin-orbit interaction (SOI).
%ADD
Furthermore, each transition metal shows huge and positive $d$-orbital 
Hall conductivity (OHC), independently of the strength of the spin-orbit interaction (SOI).
%We present an intuitive (semiclassical) explanation for the origin of the OHC,
%and explain the interpretation that orbital Hall current induces the 
%positive (negative) SHC for $n_d\gtrsim5$ ($n_d\lesssim5$) 
%in the presence of the SOI.
%discuss the close relation between the SHC and OHC.
Since the OHC is much larger than the SHC, it will be possible to realize
a {\it orbitronics device} made of transition metals.

\end{abstract}
\sloppy


\maketitle

%%%%%%%%%%%%%%%%%%%%%
%Introduction
%%%%%%%%%%%%%%%%%%%%%
\section{\label{sec:level1} INTRODUCTION}
%\subsection{\label{subsec:level1-1} Motivation and Purpose of the Study}

In recent years, spin Hall effect (SHE) in transition metals has received 
considerable attention due to its fundamental 
as well as technological interest. 
The SHE is the phenomenon that an electric field induces a 
spin current in a transverse direction.
Recent experimental efforts have revealed that many metallic compounds
show sizable spin Hall conductivity (SHC)
 \cite{Saitoh,Valenzuela,ZnSe,Kimura}.
In particular, Pt shows a huge SHC at room temperature \cite{Saitoh,Kimura}.
The observed SHC in Pt is 
$\sim240\ \hbar e^{-1} \cdot \Omega^{-1}{\rm cm}^{-1}$ \cite{Kimura}, 
which is about $10^4$ times larger than the SHC reported 
in n-type semiconductors.
%Moreover, SHC in Au also takes a large value \cite{Otani-private}.
This unexpected experimental fact cannot be understood based on the 
%semiconductor models with small Fermi surfaces.
simple electron gas models for semiconductors \cite{Murakami-SHE,Sinova-SHE}.
To elucidate the origin of the huge SHE in transition metals,
several authors have studied the SHC based on the multiorbital 
tight-binding models \cite{Kontani-Ru,Kontani-Pt}
and the first-principles band calculation \cite{Guo-Pt}.
Reference \cite{Kontani-Ru} has shown that the orbital degrees of 
freedom in transition ions, which are absent in electron gas models,
are crucial to realize the huge SHE in various transition metals.

The SHE has a close relationship to the anomalous Hall effect (AHE) 
in the presence of the magnetization ${\bf M}$,
where charge current is induced by an electric field ${\bf E}$
in parallel to ${\bf M}\times{\bf E}$.
In 1954, Karplus and Luttinger (KL) \cite{KL} solved the 
kinetic equation for $\uparrow$-, $\downarrow$-spin electrons 
%($\uparrow$-spin and $\downarrow$-spin correspond to $s_z=+1/2$ and $-1/2$,
%respectively)
in the multiband system with the $z$-component of the atomic spin-orbit
interaction (SOI) $\ld_{1}\sum_l({\hat l}_z{\hat s}_z)_l$.
Under the electric field ${\bf E}\parallel {\hat y}$,
they showed that the $\uparrow$-spin and $\downarrow$-spin electrons 
move to the opposite direction in parallel to the ${\hat x}$-axis.
Their analysis strongly suggests that the AHE (SHE) occurs
in ferromagnetic (paramagnetic) multiband systems with the SOI.
The Hall effect studied by KL, which is due to the interband
particle-hole excitation and is independent of impurity scattering,
is called the ``intrinsic Hall effect".
However, this explanation according to the KL-theory
is too naive in that they omitted the $x,y$-components of the SOI
$\lambda_2\sum_l({\hat l}_x{\hat s}_x + {\hat l}_y{\hat s}_y)_l$.
In the case of $\lambda_2\ne0$, there is no simple relation between SHE and AHE 
since $s_z=\pm1/2$ is not a good quantum number.

Note that KL did not make a mention of the SHC. The analogous relationship between AHE and SHE was first pointed out by Dyakonov and Perel \cite{Dyakonov}.

After KL, theories of the intrinsic AHE 
\cite{KL2,Fukuyama,Sinitsyn2,Kontani94,Kontani97,Miyazawa,Niu,MOnoda,Inoue-AHE,Fe,SrRuO3,Kontani06}
and the intrinsic SHE 
\cite{Murakami-SHE,Sinova-SHE,Inoue-SHE,Raimondi,Rashba,Kontani-Ru,Kontani-Pt} 
have been improved based on several specific theoretical models.
Kontani and Yamada \cite{Kontani94} studied the intrinsic AHE based on the periodic
Anderson model by considering the SOI unperturbatively. Using the microscopic
Fermi liquid theory, they derived the general expression for the anomalous Hall conductivity (AHC) by considering all the self-energy correction and the current vertex correction (CVC).
Their study had clarified that the intrinsic AHE (due to the KL mechanism)
remains finite even if all the scattering processes
are taken into account rigorously, against the Smit's claim \cite{Smit-comment}. 
The obtained general expression has succeeded in explaining the huge AHC observed in heavy-fermion systems.
It was found that the large anomalous velocity, which is not perpendicular
to the Fermi surface and is the origin of the AHE, 
is caused by $\bk$-derivative of the phase factor in $c$-$f$ mixing potential. That is, the $f$-orbital degree of freedom is significant for the AHE.
Later, AHE in $d$-electron systems has been studied intensively
 \cite{Miyazawa,Fe,SrRuO3,Kontani06}.

Recently, Murakami et al. \cite{Murakami-SHE} and Sinova et al.
 \cite{Sinova-SHE} calculated the intrinsic SHC in the Luttinger model
and the two-dimensional Rashba model, respectively.
%Consistent with their predictions, two groups \cite{Kato,Wunderlich} 
%reported optical detection of spin accumulation 
%at the sample edges in current-biased nonmagnetic semiconductors.
Later, several authors have studied the disorder effect on SHC \cite{{Inoue-SHE},{Murakami-disorder},{Bernevig}}.
Inoue et al. \cite{Inoue-AHE} proved that
the intrinsic SHE in the Rashba model vanishes due to 
the cancellation by the CVC due to impurities.
In analogy to the quantum charge Hall effect,
it has been predicted that large (and quantized) SHC may be realized
in massless Dirac electron systems,
when the chemical potential lies inside the gap induced by the SOI.
This mechanism has been predicted to be realized in
some semiconductors \cite{Murakami-qSHC} and in graphene \cite{Kane}.

However, in usual metallic systems, existence of the Dirac point
%(Dirac cone type dispersion)
just at the Fermi level cannot be expected in general.
Therefore, other novel mechanism for large SHC is expected to 
be realized in Pt and other transiton metals.
In fact, authors in ref. \cite{Kontani-Ru} presented the first report on the theoretical study of SHE in transition metal compound Sr$_2$RuO$_4$, which has no
Dirac cone type dispersion at the Fermi level.
They had found that the origin of the huge Hall effect is the ``effective Aharonov-Bohm (AB) phase" induced by $d$-orbital degrees of freedom, 
which are absent in semiconductors and in graphene.
This mechanim is expected to produce huge SHE in various 
multiorbital $d$-electron systems universally.
%Kontani et al. \cite{Kontani-Ru} showed that the anomalous 
%velocity due to the $d$-orbital degrees of freedom causes the 
%arge SHE in transition metals universally.
Later, Ref. \cite{Kontani-Pt,Guo-Pt} succeeded in reproducing the SHC in Pt 
theoretically. %, by taking account of the realistic band structure.

In the present paper,
we study the intrinsic SHE in various $4d$- and $5d$-transition metals
by taking account of their realistic band structures.
We employ the Naval Research Laboratory tight-binding (NRL-TB) 
model \cite{NRL1,NRL2}, which enables us to construct 
nine-orbital ($s+p+d$) tight binding models for each transition metal.
We find that both Pt ($5d^9$) and Pd ($4d^9$), which have face-centered 
cubic (fcc) structure, show large positive SHCs.
On the other hand, the SHCs take large negative values 
in Ta ($5d^4$), Nb ($4d^4$), W ($5d^5$), and Mo ($4d^5$), 
which have body-centered cubic (bcc) structures.
We find that the SHC changes smoothly with the electron number $n=n_s + n_d$ 
regardless of the changes of the crystal structure, where $n_{\alpha}$ represents the number 
of electrons on $\alpha$-orbital.
Among them, Pt shows the largest SHC in magnitude in the low resistive regime.
%The obtained SHC is sensitive to the changes in the chemical potential 
%$\mu$, reflecting the characteristic feature of the band structure. 
In usual, intrinsic SHC is independent of resistivity in the low
resistive regime ($\rho \lesssim 50 \ \mu\Omega{\rm cm}$), whereas
it decreases in proportion to $\rho^{-2}$  in the high resistive regime.
However, we find a condition that the intrinsic SHC decreases
as $\rho$ approaches zero in the low resistive regime.
This anomalous phenomenon may be realized in Ta.

%ADD
Furthermore, we study the $d$-orbital Hall effect (OHE), 
which is the phenomenon that an electric field induces a 
$d$-orbital current in a transverse direction \cite{Kontani-Ru,Kontani-Pt}.
We find that the $d$-orbital Hall conductivity (OHC) is 
almost one order of magnitude larger than the SHC,
since the OHC occurs even in the absence of the SOI.
Using the large OHE in transition metals, we will be able to constract a 
{\it orbitronics device} made of transition metals.
In a later publication, we will present an intuitive (semiclassical) 
explanation for the origin of the OHC \cite{future}.
%In \S \ref{subsec:level5-2}, we present an intuitive (semiclassical) 
%explanation for the origin of the OHC.
%We can consider that the OHE is the essential phenomenon,
%and the SHE is a passive phenomenon induced by the OHE if SOI presents.
%In this sence, the OHE is the origin of the SHE and the AHE in transition metals.

%Here, we present an intuitive explaination for the origin of the intrinsic Hall effect after refs. \cite{Kontani-Ru,Kontani-Pt}. In multiorbital systems, a conduction electron acquires the trajectory-dependent phase factor that originates from the atomic wavefunction. This phase factor can be interpreted as the AB phase due to the ``effective magnetic flux'' that is caused by the angular dependence of the atomic wavefunction with the aid of the SOI.
%Since the effective magnetic flux for the $\uparrow$-spin electron and that for the $\downarrow$-spin electron are opposite in sign, $\uparrow$-spin and $\downarrow$-spin electrons move separately. Therfore, effective magnetic flux that is inherent in the multiorbital systems is the origin of the SHE in paramagnetic metals \cite{Kontani-Ru,Kontani-Pt}, and that of the AHE in ferromagnetic metals \cite{Kontani94}.

Finally, we comment on the extrinsic Hall effect.
In 1958, Smit \cite{Smit} studied the AHE due to the asymmetric scattering 
around the impurity in the presence of spin-orbit coupling, which is called
the skew-scattering mechanism.
The AHC due to the skew-scattering is 
proportional to $\rho^{-1}$ if elastic scattering is dominant.
In 1970, Berger proposed another mechanism of extrinsic Hall effect,
the side jump due to impurities \cite{Berger}.
This mechanism gives the AHC in proportion to $\rho^{-2}$.
Both extrinsic Hall effects vanish where the inelastic scattering due to electron-electron or electron-phonon interaction is dominant over the elastic scattering.
Both mechanisms (the skew-scattering and the side jump)
cause the extrinsic SHE \cite{dyakonov,hirsh,zhang,Halperin}.
In the present paper, we do not study the extrinsic SHE,
which is sensitive to the character of the impurity potential.
It is an important future problem to study the extrinsic SHE
in realistic multiorbital tight-binding models.

%%%%%%%%%%%%%%%%%%%%%
%Model and Hamiltonian
%%%%%%%%%%%%%%%%%%%%%
\section{\label{sec:level2} MODEL AND HAMILTONIAN}

%In the pesent study, we use the Naval Research Laboratory tight-binding (NRL-TB) model \cite{{NRL1},{NRL2}} to obtain the bandstructure in various metals. The NRL-TB model employs the scheme of two-center, non-orthognal Slater-Koster (SK) Hamiltonian \cite{SK}. SK parameters are represented with distance- and environment dependent parameters that are determined so that the total energy and the band structures agree well with those obtained with first-principles calculations. To describe the electronic state in various metals, we consider 5s,5p,4d or 6s,6p,5d orbitals (nine orbital per atom are considered) and hopping integrals between up to sixth nearest neighbor sites. The NRL-model uses the non-orthognal basis. 

In the present study, we use the NRL-TB model \cite{{NRL1},{NRL2}} to obtain the band structure in various transition metals. Here, we shortly explain this model.
The NRL-TB model employs the scheme of two-center and non-orthogonal Slater-Koster (SK) Hamiltonian \cite{SK}. The SK parameters are represented with distance- and environment dependent parameters that are determined so that the total energy and the band structures agree with those obtained by full-potential LAPW LDA calculations. 
The root mean square error in the fitting is about 0.002-0.004 Ry \cite{Papas}. 
This fitting error is small enough to perform reliable numerical calculations.
To describe the electronic
state in 4$d$ (5$d$) metals, we consider 5$s$, 5$p$, and 4$d$ (6$s$, 6$p$, and 5$d$) orbitals, that is, we consider nine orbitals per atom. Hopping integrals between a pair of atoms are then expressed with ten SK parameters 
($ss\sigma$, $sp\sigma$,
$pp\sigma$, $pp\pi$, $sd\sigma$, $pd\sigma$, $pd\pi$, $dd\sigma$,
$dd\pi$, and $dd\delta$). 
In this study, we consider hopping (and overlap) integrals between up to sixth nearest neighbor sites for metals with fcc and bcc structures, and ninth nearest neighbor sites for hexagonal closed packed
(hcp) structures. 

Table \ref{table1} shows the crystal structure, electron number per atom, and the coupling constant $\ld$ of SOI $\ld\sum_{i} \hat l_i \hat s_i$ $(i=x, y, z)$ for various 4$d$- and 5$d$- transition metals. $(m_d d^{n_d} m_s s^{n_s})$ represent the electronic configuration of an isolated atom, where $m_d$ and $m_s$ is the main quantum number, and $n_s$ and $n_d$ is the number of electrons on $s$- and $d$-orbital. In this table, we put $\ld$=0.03 Ry for 5$d$ electron in Pt and $\ld$=0.013 Ry for 4$d$ electron in Pd, according to optical spectroscopy \cite{Friedel}. 
For other 4$d$- and 5$d$- transition metals, we used Herman-Skillman atomic spin-orbit parameters \cite{Herman}: These parameters had been calculated by using the self-consistent Hartree-Fock-Slater atomic functions. Here, we consider only the $d$-orbital SOI, and neglect other SOI terms which may possess a $\bk$-dependence.
Hereafter, we set the unit of energy Ry; 1 Ry = 13.6 eV. 
\begin{table}[!htb]
\caption{\label{table1} The crystal structure, electron number per atom, and the coupling constant $\ld$ of SOI for various transition metals. $(m_d d^{n_d} m_s s^{n_s})$ represent the electronic configration of an isolated atom. Here, $m_d$ and $m_s$ is the main quantum number, and $n_s$ and $n_d$ is the number of electrons on $s$- and $d$-orbital. Bcc, hcp and fcc represents a body-centered cubic, hexagonal closed packed and face-centered cubic, respectively.}
\begin{ruledtabular}
\begin{tabular}{lrrr}
metals & structure & electron number & SOI (Ry)\\ \hline
Nb & bcc  &5 (4$d^4$5$s^1$)&  0.006 \\
Mo & bcc &6 (4$d^5$5$s^1$)& 0.007 \\
Tc & hcp & 7 (4$d^6$5$s^1$) & 0.009 \\
Ru & hcp & 8 (4$d^7$5$s^1$) & 0.01 \\
Rh &  fcc &9 (4$d^8$5$s^1$)& 0.011 \\
Pd &  fcc  &10 (4$d^{10}$5$s^0$)& 0.013\\
Ag & fcc & 11 (4$d^{10}$5$s^1$) & 0.019 \\ \hline
Ta &  bcc &5 (5$d^3$6$s^2$)&  0.023 \\
W & bcc  &6 (5$d^4$6$s^2$)& 0.027 \\
Re & hcp & 7 (5$d^5$6$s^2$) & 0.025 \\
Os & hcp & 8 (5$d^6$6$s^2$) & 0.025 \\
Ir  &  fcc  &9 (5$d^9$6$s^0$)& 0.025 \\
Pt & fcc  & 10 (5$d^9$6$s^1$)& 0.03 \\
Au & fcc & 11 (5$d^{10}$6$s^1$) & 0.03
\end{tabular}
\end{ruledtabular}
\end{table}

In the presence of SOI for 4$d$ or 5$d$ electrons, the total Hamiltonian is given by 
\begin{\eq}
\hat H =
\left(
\begin{array}{cc}
\hat H_0 + \ld \hat l_z/2 & \ld (\hat l_x - i \hat l_y)/2 \\
\ld (\hat l_x + i \hat l_y)/2 & \hat H_0 - \ld \hat l_z/2
\end{array}
\right),
\end{\eq}
where the first and the second rows (columns) correspond to $s_z=+\hbar/2$ ($\uparrow$-spin) and $s_z=-\hbar/2$ ($\downarrow$-spin). $\hat H_0$ is a $9\times 9$ matrix given by NRL-TB model for bcc and fcc structure. In the case of hcp structure, $\hat H_0$ is an 18$\times$18 matrix, since a unit cell contains two atoms. 
The matrix elements of $\bm{l}$ for $d$-orbital are given by \cite{Friedel}
\begin{\eq}
l_x&=\left(
\begin{array}{ccccc}
0 & 0 & -i &0 &0 \\
0 & 0& 0& -i & -i\sqrt{3} \\
i & 0 & 0& 0 & 0 \\
0& i& 0& 0& 0 \\
0& i\sqrt{3} & 0& 0& 0
\end{array}
\right), \nn
\end{\eq}
\begin{\eq}
l_y&=\left(
\begin{array}{ccccc}
0& i& 0& 0& 0 \\
-i& 0& 0& 0& 0 \\
0& 0& 0& -i & i\sqrt{3} \\
0& 0& i& 0 & 0 \\
0& 0& -i\sqrt{3} &0 & 0 
\end{array}
\right), \nn
\end{\eq}
\begin{\eq}
l_z&=\left(
\begin{array}{ccccc}
0 &0 &0 &2i &0 \\
0 &0 &i &0  &0 \\
0& -i& 0& 0& 0 \\
-2i& 0& 0& 0& 0 \\
0& 0& 0& 0& 0  
\end{array}
\right).
\end{\eq} 
where the first to the fifth rows (columns) correspond to $d$-orbitals $xy$, $yz$, $zx$, $x^2-y^2$, and $3z^2-r^2$, respectively.

In NRL-TB model, we use the non-orthogonal basis since the atomic wave functions of different sites are not orthogonal:
\begin{\eq}
\int d\br \phi^{\ast }_{\alpha} (\br - \bR_{i}) \phi_{\beta}(\br -\bR_{i'}) = O_{\alpha\beta}(\bR_{i}-\bR_{i'}), \label{eq:o1}
\end{\eq}
where $\phi_{\alpha}(\br - \bR_i)$ represents the atomic wave function at the $i$th site and $\alpha,\beta$ is orbital state indices, and $O_{\alpha\beta}(\bR_{i}-\bR_{i'})$ represents the overlap integral between different sites. When the overlap integrals between different sites are negligible, eq. (\ref{eq:o1}) is simplified as 
\begin{\eq}
O_{\alpha\beta}(\bR_{i}-\bR_{i'}) = \delta_{\alpha\beta}\delta_{ii'}. \label{eq:o0}
\end{\eq}
This approximation is rather appropriate when $\alpha$ and $\beta$ correspond to $d$-orbitals, since $d$-orbital atomic wave functions are localized well. However, when $\alpha$ and $\beta$ are either $s$- or $p$-orbital, $O_{\alpha\beta}(\bR_{i}-\bR_{i'})$ is large even when $i\neq i'$, and therefore eq. (\ref{eq:o0}) is not satisfied.

\begin{figure}[!htb]
\includegraphics[width=.75\linewidth]{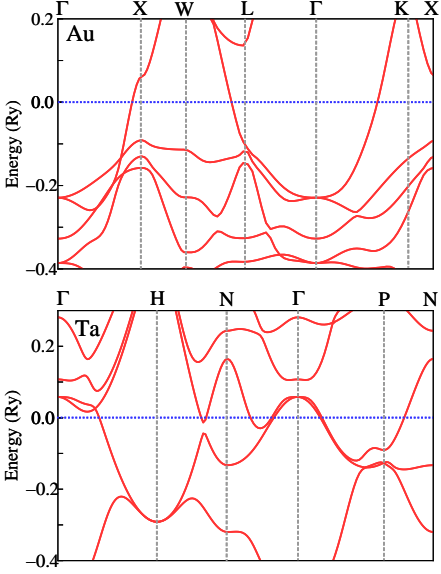}
\caption{\label{fig:dispersion} (upper panel) Band structure of Au. Here, $\Gamma=(0,0,0)$, X=$(\pi,0,0)$, W=$(\pi,\pi/2,0)$, L=$(\pi/2,\pi/2,\pi/2)$, and K=$(3\pi/4,3\pi/4,0)$. (lower panel) Band structure of Ta. Here, $\Gamma=(0,0,0)$, H=$(\pi,0,0)$, N=$(\pi/2,\pi/2,0)$, P=$(\pi/2,\pi/2,\pi/2)$. 
%add
Near the Fermi level, we see that the band structures obtained in the present model agree well with the result of the relativistic first-principles calculation in refs. \cite{{Mattheiss},{Kellen}}.
}
\end{figure}

The band structures obtained for the present model in Au and Ta are shown in Fig. \ref{fig:dispersion}.
These band structures are derived by taking the overlap integrals between different sites into account correctly. The methods how to calculate these band structures are explained in section \ref{subsec:level3-3}.
Near the Fermi level, the obtained band structures are in good agreement with the results of the relativistic first-principles calculations \cite{{Mattheiss},{Kellen}}.
Since the Fermi surface is mainly composed of $d$-electrons in the transition metals, in which we study the SHE and OHE, the band structure near the Fermi level is described well when $O_{\alpha\beta}(\bR_{i}-\bR_{i'})$ is approximated by eq. (\ref{eq:o0}). Therefore, the calculations of SHE and OHE using eq. (\ref{eq:o0}) seems to give semiquantitatively reliable results \cite{Kontani-Pt}.
However, for a more quantitative study of SHE and OHE, we need to consider the overlap integrals between different sites.

%%%%%%%%%%%%%%%%%%%%%%%%%%%%%%%%%
%%%%%%%%%%%%%%%%%%%%%%%%%%%%%%%%%

Until section \ref{subsec:level3-3}, we will use the simplified overlap integrals given by eq. (\ref{eq:o0}) to simplify the explanation. In section \ref{subsec:level3-3}, we will study the SHE and OHE by considering the overlap integrals in eq. (\ref{eq:o1}) correctly.
The 18$\times$18 matrix form of the Green function without impurities is given by $\hat G^0(\bk,\w)=(\w+\mu-\hat H )^{-1}$
where $\mu$ represents the chemical potential. There is a $\bk$-dependent unitary matrix $\hat U$ which diagonalize the Hamiltonian $\hat H$ as follows:
\begin{\eq}
\sum_{\alpha\beta} U^{\dagger}_{l\alpha} H^{\alpha\beta} U_{\beta m} = E^l_{\bk} \delta_{lm}, \label{eq:U}
\end{\eq} 
where $\alpha,\beta$ is the orbital indices, and $l$ and $m$ are the band indices. 

Here, we consider the quasiparticle damping rate $\hat \Gamma$, which is given by $ \ldk \hat \Sigma_{\bk}(-i0) - \hat \Sigma_{\bk}(+i0) \rdk/2i$, where $\hat \Sigma_{\bk}(\w)$ is the self-energy matrix. 
In the Born approximation for $I \ll W_{band}$, $\hat \Gamma$ is given by $\hat \Gamma=n_{imp} I^2 \f{1}{2i} \ldk \hat g(-i0) - \hat g(+i0) \rdk$, where $n_{imp}$ is the impurity concentration, $I$ is the impurity potential, $W_{band}$ is the bandwidth, and $\hat g(\w)$ is the local Green function $\hat g(\w)= \f{1}{N} \sum_{\bk} \hat G(\bk,\w)$, respectively. Here, $N$ is the total number of lattice points. In the $T$-matrix approximation for general strengh of $I$, $\hat \Gamma$ is given by $\displaystyle \hat \Gamma = n_{imp} I \f{1}{2i} \ltk \f{1}{1-I\hat g(-i0)}-\f{1}{1-I\hat g(+i0)} \rtk$. The retarded and advanced Green functions are given by
\begin{\eq}
\hat G^R(\bk,\w) &= (\w+\mu-\hat H +  i \hat \Gamma)^{-1}, \\
\hat G^A(\bk,\w) &= (\w+\mu-\hat H -  i \hat \Gamma)^{-1}.
\end{\eq}
%For simplicity, we assume that $\hat \Gamma$ is diagonal with respect to orbital, and independent of momentum. Then, the retarded Green function is given by 
%This assumption will be justified when the damping rate is caused by local impurities since the local Green function $g_{\alpha\beta}(\w)$=$\f{1}{N}\sum_{\bk} G_{\alpha\beta}(\bk,\w)$ is diagonal and independent of ($\alpha,\beta$) if $\ld=0$ \cite{Kontani06}. Here, $\alpha,\beta$ represent orbital indices, and $N$ is the total number of lattice points, respectively.

In the present model, the charge current operator for $\mu$-direction ($\mu=x,y$) is given by
\begin{\eq}
\hat J^C_{\mu}=
\left(
\begin{array}{cc}
\hat j_{\mu}^C & 0 \\
0 & \hat j_{\mu}^C
\end{array}
\right),  \label{eq:chargecurrent}
\end{\eq}
where $\hat j_{\mu}^C = -e \f{\rd \hat H_0}{\rd k_{\mu}}$, and $-e$ $(e>0)$ is the electron charge. Here, atomic SOI is not involved in the charge current operator since it is $\bk$-independent. Also, the $s_z$-spin current operator $\hat J^S_{\mu}= \ltk \hat J^C_{\mu}, \hat s_z \rtk/2 = \left( \hat J^C_{\mu}\hat s^z + \hat s^z \hat J^C_{\mu} \right)/2$ \cite{{Inoue-SHE},{Guo}} is given by 
\begin{\eq}
\hat J^S_{\mu}=
(-\hbar/e)\left(
\begin{array}{cc}
\hat j^C_{\mu} & 0 \\
0 & - \hat j^{C}_{\mu}  
\end{array}
\right), \label{eq:Js}
\end{\eq}
and $l_z$-orbital current operator is given by
\begin{\eq}
\hat J^O_x=\ltk \hat J^C_x, \hat l_z \rtk/2. \label{eq:Jo}
\end{\eq}
We will discuss the validity of these current operators in Appendix \ref{level:App-C} in more detail.

%%%%%%%%%%%%%%%%%%%%%%%%%%%%
% Derivation of SHC and OHC
%%%%%%%%%%%%%%%%%%%%%%%%%%%%

\section{\label{sec:level3} SPIN and ORBITAL HALL CONDUCTIVITIES}
\subsection{\label{subsec:level3-1} SHC and OHC without overlap integrals between different sites}
In this section, we derive the general expressions for the intrinsic SHE and OHE based on the linear-response theory. 
As we will discuss in \ref{subsec:level3-2}, we can safely neglect the CVC in calculating SHC and OHC in the present model. Therefore, the SHC at T=0 is given by $\sxy^z = \sxy^{zI} + \sxy^{zII}$ according to Streda \cite{Streda}, where
\begin{\eq}
\sxy^{zI}&=& \f{1}{2\pi N}\sum_{\bk} \text{Tr} \left[ \hat J^S_x \hat G^R \hat J^C_y \hat G^A \right]_{\w =0} , \label{eq:sxyI} \\
\sxy^{zII}&=& \f{-1}{4\pi N} \sum_{\bk} \int_{-\infty}^{0} d\w \text{Tr} \left[  \hat J^S_x \f{\partial \hat G^R}{\partial \w}\hat J^C_y \hat G^R \right. \nn
&&\left.  - \hat J^S_x \hat G^R \hat J^C_y \f{\partial \hat G^R}{\partial \w} - \langle R \leftrightarrow A \rangle \right]. \label{eq:sxyII}
\end{\eq}
Here, $I$ and $II$ represent the ``Fermi surface term" and the ``Fermi sea term", respectively. In the same way, the OHC of the Fermi surface term $O^{zI}_{xy}$ and that of the Fermi sea term $O^{zII}_{xy}$ are respectively given by eqs. (\ref{eq:sxyI}) and (\ref{eq:sxyII}) by replacing $\hat J^{S}_x$ with the $l_z$-orbital current operator $\hat J^{O}_x$ in eq. (\ref{eq:Jo}).

In the Born approximation, the quasiparticle damping rate depends on orbital indices. If $\ld=0$, $\hat \Gamma$ is diagonal with respect to orbital: $\hat \Gamma_{\alpha\beta} = \g_{\alpha} \delta_{\alpha\beta}$, where $\alpha$ and $\beta$ are orbital indices. Then, the offdiagonal terms are negligible when $\ld \ll W_{band}$. $\g_{\alpha}$ is the quasiparticle damping rate for $\alpha$-orbital, and is proportional to the local density of states (LDOS) for $\alpha$-orbital, $\rho_{\alpha}(0)$.
In the present model, $d$-orbital LDOSs are about equal in magnitude. Therefore, $\g_{\alpha}$ is approximately independent of $\alpha$ and can be approximated by a constant $\g$: $\Gamma_{\alpha\beta} = \g\delta_{\alpha\beta}$. 
Using this constant $\g$ approximation, we derive the general expressions for the Fermi surface term and the Fermi sea term. In this case, the retarded and advanced Green function can be diagonalized using unitary matrix $\hat U$ given by eq. (\ref{eq:U}) as follows:
\begin{\eq}
\sum_{\alpha\beta} U^{\dagger}_{l\alpha} G^{R}_{\alpha\beta} U_{\beta m} = \f{\delta_{lm}}{\w - E^{l}_{\bk} + i\g},
\end{\eq}
where $l,m$ is the band indicies.
Therefore, we can rewrite eqs. (\ref{eq:sxyI}) and (\ref{eq:sxyII}) by using $\hat U$ as follows: 
\begin{\eq}
\sxy^{zI} &=& \f{1}{2\pi N} \sum_{\bk,l \neq m} (J^S_x)^{ml} (J^C_y)^{lm} \f{1}{(E^l_{\bk}-i\g)(E^m_{\bk}+i\g)} \nn
 &=& \f{-1}{2\pi N} \sum_{\bk,l \neq m} \text{Im} \ltk (J^S_x)^{ml} (J^C_y)^{lm} \rtk \nn 
&& \times \text{Im} \ldk \f{1}{(E^l_{\bk}-i\g)(E^m_{\bk}+i\g)} \rdk, \label{eq:sxyI-1}
\end{\eq}
\begin{\eq}
\sxy^{zII} &=& -\f{1}{2\pi N} \sum_{\bk, l\neq m} \int_{-\infty}^{0} d\w \text{Im} \ltk (J^S_x)^{ml} (J^C_y)^{lm} \rtk \nn
&&\times \text{Im} \ldk \f{1}{(\w- E^l_{\bk} +i\g)^2} \f{1}{(\w-E^m_{\bk} + i\g)} \right. \nn
&&\left. -\f{1}{(\w-E^l_{\bk}+i\g)} \f{1}{(\w-E^m_{\bk} +i\g)^2}\rdk, \label{eq:sxyII-1}
\end{\eq}
where $(J^S_x)^{ml}$ is given by $\sum_{\alpha\beta} U^{\dagger}_{m\alpha} (J^S_x)^{\alpha\beta} U_{\beta l} $. Note that we dropped the diagonal terms $l=m$ in the summations in eqs. (\ref{eq:sxyI-1}) and (\ref{eq:sxyII-1}) since they vanish identically. We also note that the transformation from the first row to the second row in eq. (\ref{eq:sxyI-1}) was performed since $\sum_{l,m} \text{Re} \ltk (J^S_x)^{ml}(J^C_y)^{lm} \rtk$ vanishes identically after $\bk$-summation.
After performing the $\w$-integration in eq. (\ref{eq:sxyII-1}), the Fermi sea term is given by $\sxy^{zII} = \sxy^{zIIa} + \sxy^{zIIb}$, where
\begin{align}
\sxy^{zIIa}= \f{-1}{2\pi N} &\sum_{\bk, l\neq m} \text{Im} \ltk (J^S_x)^{ml}(J^C_y)^{lm} \rtk \f{1}{E^l_{\bk}-E^m_{\bk}} \nn
&\times \text{Im} \ltk \f{E^l_{\bk} + E^m_{\bk}- 2i\g}{(E^l_{\bk}-i\g )(E^m_{\bk}-i\g )} \rtk, \label{eq:IIa} \\
\sxy^{zIIb}= \f{1}{\pi N} &\sum_{\bk, l\neq m} \text{Im} \ltk (J^S_x)^{ml}(J^C_y)^{lm} \rtk\f{1}{(E^l_{\bk}-E^m_{\bk})^2} \nn 
&\times \text{Im}\ltk \ln \left( \f{E^l_{\bk}-i\g}{E^m_{\bk}-i\g} \right)\rtk. \label{eq:IIb}
\end{align}
Here, we used the following relation to perform the $\w$-integration: 
\begin{align}
&\int_{-\infty}^{\mu} dx \ltk \f{1}{(x-a)^2(x-b)} -\f{1}{(x-a)(x-b)^2} \rtk = \nn
&\f{a+b-2\mu}{(a-b)(a-\mu)(b-\mu)} -\f{2}{(a-b)^2} \text{ln}\lk \f{a-\mu}{b-\mu} \rk.
\end{align}

In the case of $\g\rightarrow 0$, $\sxy^{zIIb}$ given by eq. (\ref{eq:IIb}) corresponds to the Berry cuvature term given by:
\begin{\eq}
\sxy^{zIIb} = \f{1}{N} \sum_{\bk,l} f(E^l_{\bk})\Omega^l(\bk),
\end{\eq}
where $\Omega^l(\bk)$ represents the Berry curvature given by
\begin{\eq}
\Omega^l(\bk) = \sum_{m\neq l} \f{2 \text{Im} \ltk \left( J^S_x \right)^{ml}\left( J^C_y \right)^{lm}  \rtk}{(E^l_{\bk}-E^m_{\bk})^2}.
\end{\eq}
The relation $\sxy^z=\sxy^{zIIb}$ has been frequently assumed in literatures such as ref. \cite{Guo-Pt}.
In this study, we calculate all the Fermi surface and Fermi sea term correctly, and elucidate how each term contribute to the SHC and OHC.
In section \ref{subsec:level5-3}, we will discuss the $\g$-dependece of $\sxy^{zI}, \sxy^{zIIa}$ and $\sxy^{zIIb}$ in detail.

%%%%%%%%%%%%%%%%%%%%%%%%%%
%Discussion of the CVC
%%%%%%%%%%%%%%%%%%%%%%%%%%

\subsection{\label{subsec:level3-2} Discussion on the CVC}
\begin{figure}[!htp]
\includegraphics[width=.8\linewidth,height=.2\linewidth]{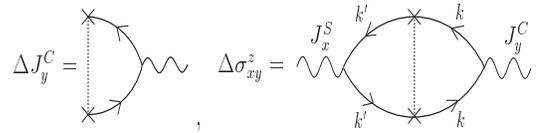}
\caption{\label{fig:CVC} The diagrammatic expressions of the current vertex correction due to the local impurity potentials. Here, the diagram represents the lowest order correction to the current $\Delta \hat J^C_x$ and spin Hall conductivity $\Delta \sxy^{z}$.} 
\end{figure}

In section \ref{subsec:level3-1}, we have neglected the CVC.
Here, we calculate the CVC due to the local impurity potential in the Born approximation, and show that it is negligible in transition metals. In the Born approximation, the lowest order CVC is given by 
\begin{\eq}
\Delta \hat J^C_{\mu} = \f{1}{N} n_{imp} I^2 \sum_{\bk} \hat G^R \hat J^C_{\mu} \hat G^A. \label{eq:deltaJ}
\end{\eq}
Its diagrammatic expression is given in Fig. \ref{fig:CVC}. The magnitude of CVC depends on the model. For example, the CVC vanishes identically in the $d$-orbital tight-binding models with atomic SOI \cite{{Kontani-Ru},{Kontani06}}. In contrast, the CVC plays an essensial role in a Rashba model: the SHC vanishes due to the cancellation by CVC 
\cite{Inoue-AHE}. 

%rewrite
Here, we study the CVC in fcc and bcc transition metals, where each atomic site is a center of 
inversion symmetry.
The $s$- and $d$- orbital atomic wave functions have an even parity with respect to $\bk\rightarrow -\bk$, whereas the $p$-orbital atomic wave functions have an odd parity. 
Therefore, the Hamiltonian for a fcc and bcc metal has a following relationship:
\begin{\eq}
(H_0(\bk))_{\alpha\beta} = p (H_0(-\bk))_{\alpha\beta},
\end{\eq}
where $p=-1$ only when either $\alpha$ or $\beta$ is $p$-orbital, otherwise $p=1$. It is easy to show that $(G(\bk,\w))_{\alpha\beta}$ has the same parity with $(H_0(\bk))_{\alpha\beta}$. Therefore, when both $\alpha$ and $\beta$ are $(s,d)$- orbitals, $(\Delta \hat J^C_{\mu})_{\alpha\beta}=0$ since the $(\alpha,\beta)$-components of $(\partial/\partial k_{\mu}) \hat G= \hat G \hat J^C_{\mu} \hat G$ is an odd function. On the other hand, $(\Delta \hat J^C_{\mu})_{\alpha\beta}\neq 0$ when either $\alpha$ or $\beta$ is $p$-orbital since $(\partial \hat G /\partial k_{\mu})_{\alpha\beta}$ is an even function. Although $(\Delta \hat J^C_{\mu})_{\alpha\beta}$ originating from the $p$-orbital is finite, it is small in magnitude since the $5p,6p$-level is about 20eV higher than the Fermi level $\mu$ and the $p$-electron density of states (DOS) at $\mu$ is very small in all transition metals.

%$(\hat H_0(\bk))_{\alpha,\beta}$ is an odd function with respect to $\bk \leftrightarrow  -\bk$ when $\alpha$ is one of $p$-orbitals and $\beta$ is one of $(s,d)$- orbitals. It is easy to show that $(G(\bk,\w))_{\alpha\beta}$ and $(H_0(\bk))_{\alpha\beta}$ have the same parity with respect to $\bk\rightarrow -\bk$.
%Therefore, $(\Delta \hat J^C_{\mu})_{\alpha\beta}$ is finite when either $\alpha$ or $\beta$ is $p$-orbital since the $(\alpha,\beta)$-component of $(\partial/\partial k_{\mu}) \hat G= \hat G \hat J^C_{\mu} \hat G$ is an even function. However, the CVC that originates from $p$-orbital is small since the $5p,6p$-level is about 20eV higher than the Fermi level $\mu$ and the $p$-electron density of states (DOS) at $\mu$ is very small in all transition metals. 
%%%%%%%%%%%%%%%%%%%%%%%%%%
%%%%%%%%%%%%%%%%%%%%%%%%%%%

Now, let us verify numerically that the contribution of CVC to SHC is small in magnitude.
The correction to the SHC due to the lowest order CVC is given by
\begin{align}
\Delta \sxy^z &= \f{1}{2\pi N^2} n_{imp} I^2 \sum_{\bk,\bk'} \text{Tr} \ldk \hat J^S_x \hat G^R_{\bk'} \Delta \hat J^C_y \hat G^A_{\bk'} \rdk \nn
                   &= \f{1}{2\pi N^2} n_{imp} I^2 \sum_{\bk,\bk'} \text{Tr} \ldk \hat G^A_{\bk'} \hat J^S_x \hat G^R_{\bk'} \hat G^R_{\bk} \hat J^C_y \hat G^A_{\bk} \rdk.
\end{align}
Its diagrammatic expression is given in Fig. \ref{fig:CVC}. 
We calculated $\Delta \sxy^z$ numerically and found that it is very small compared to $\sxy^z$ without CVC: the ratio $\mid  \Delta \sxy^z \mid $ / $\mid  \sxy^z \mid $ is $\sim$ 0.02 for Ta and Pt, and $\sim$ 0.005 for W when $\g$=0.002$\sim$0.02. 
The ratio is independent of $\g$ by the following reason:
Since $\sum_{\bk} G^R_{\bk} G^A_{\bk} \sim O(\g^{-1})$, and $\g\propto n_{imp}I^2$, $\Delta J^C_{\mu}$ in eq. (\ref{eq:deltaJ}) is independent of $\g$.
In the present model, the higher order correction to the SHC should be negligible.

In the case of hcp transition metals, $(\Delta \hat J^C_{\mu})_{\alpha\beta}\neq 0$ even when
$\alpha$ and $\beta$ are ($s,d$)-orbitals, since each atomic site is not a center of inversion symmetry. 
To find out the importance of the CVC in hcp metals,
we performed the numerical calculation for Os, and found that the ratio 
$\mid  \Delta \sxy^z \mid $ / $\mid  \sxy^z \mid $ is $\sim 0.06$ in Os
in the low resistive regime.
Although it is much larger than that in Pt, Ta and W,
the CVC is qualitatively negligible even in hcp transition metals.
Therefore, we are allowed to neglect the CVC even for hcp transition metals.

%Similarly, in the $T$-matrix approximation for the general strength of $I$, the CVC is always proportional to
%\begin{\eq}
%\Delta \hat J^C_{\mu}\propto \sum_{\bk} \hat G^R_{\bk} \hat J^C_{\mu}(\bk) \hat G^A_{\bk}.
%\end{\eq}
%By considering the parity of the Green function and the charge current operator as explained above, the CVC due to local impurity vanishes identically in the $d$-orbital tight-binding model.
%Also, in the present model, the CVC which originates from $p$-orbital will be small in magnitude.
%Therefore, the CVC due to local impurity is small in magnitude in the present model, and it vanishes identically in the $d$-orbital tight-binding models.
%Therfore, CVC is independent of $n_{imp}$ and $I$, which can be safely neglected for the SHC in transition metals. 

%Thus, we can safely neglect the CVC for the SHC in transition metals.

% ADD
%In Appendix \ref{APP-D}, 
%we briefly study the vertex correction (VC) for spin current
%caused by the Coulomb interaction, which is reffered to as
%spin Coulomb drag effect \cite{Coulomb-drag}.
%This study suggests that the SHC is (quantitatively) reduced by the VC 
%for spin current in systems with strong Coulomb interaction.
%Therefore, a simple relationship $\sxy^z = -\f{\hbar}{e}\sxy$
%is not satisfied in strongly correlated systems due to spin Coulomb drag.

%%%%%%%%%%%%%%%%%%%%%%%%%%%%%%%%%
% SHC and OHC with Overlap integrals
%%%%%%%%%%%%%%%%%%%%%%%%%%%%%%%%%

\subsection{\label{subsec:level3-3} SHC and OHC considering the overlap integrals between different sites}

In the previous section, we studied the SHC and OHC under the assumption that the atomic wave functions of different sites are orthogonal.
However, for a more accurate quantitative study of the intrinsic SHE and OHE, we need to take the off-diagonal elements of $O_{\alpha\beta}$ in eq. (\ref{eq:o1}) into account correctly. 
%Although the overlap integrals of $d$-orbitals are small, 
In this subsection, we explain how to calculate the SHE and OHE when the overlap integrals between different sites are considered.

Here, we introduce the Bloch wave function which is given by 
\begin{\eq}
\phi_{\bk \alpha} (\br ) = \f{1}{\sqrt{N}} \sum_i \e^{i\bk\cdot \bR_i} \phi_{\alpha}(\br-\bR_i). \label{eq:Bloch}
\end{\eq}
In this case, the inner product between the Bloch wave functions with different $\bk$ and $\alpha$ is given by 
\begin{\eq} 
\int d\br \phi^{\ast }_{\bk \alpha} (\br ) \phi_{\bk' \beta} (\br ) = \delta_{\bk\bk'}O_{\alpha\beta}(\bk), \label{eq:o1-k}
\end{\eq}
where
\begin{\eq}
O_{\alpha\beta}(\bk) &=& \sum_i \e^{-i \bk\cdot \bR_i} O_{\alpha\beta}(\bR_i). \label{eq:Ok}
\end{\eq}
Here, $O_{\alpha\beta}(\bR_i)$ in eq. (\ref{eq:Ok}) is the overlap integral which is defined by eq. (\ref{eq:o1}).
Therefore, when the overlap integrals between different sites are considered, the Bloch wave function given by eq. (\ref{eq:Bloch}) is non-orthogonal.

By including the chemical potential $\mu$, the kinetic term of Hamiltonian is given by \cite{Naito}
\begin{\eq}
\hat H_0 = \sum_{\bk,\alpha,\beta} c^{\dagger}_{\bk \alpha} \ldk h_{\alpha\beta}(\bk) -\mu O_{\alpha\beta}(\bk) \rdk c_{\bk \beta}, 
\label{eq:ham}
\end{\eq}
where $c_{\bk \alpha}$ is defined by 
\begin{\eq}
c_{\bk \alpha }= \f{1}{\sqrt{N}} \sum_{i} \text{e}^{i \bk\cdot \bm{R_i}} c_{i \alpha}.
\end{\eq}
Here, $c_{i \alpha}$ is an annihilation opertaor of an electron in the $\alpha$ orbital state at $i$th site. As the atomic wave functions at different sites are non-orthogonal, creation and annihilation operators $c^{\dagger}_{\bk \alpha}, c_{\bk \alpha}$ do not satisfy the canonical anticommutation relations, but instead satify \cite{Naito}
\begin{\eq}
\{ c_{\bk \alpha}, c^{\dagger}_{\bk' \beta} \} = \delta_{\bk \bk'} O^{-1}_{\alpha\beta}(\bk).
\end{\eq}
Since matrix $\hat O(\bk)$ is positive definite Hermitian matrix, we can introduce a matrix $\hat S(\bk)$ that transforms $\hat O (\bk)$ into the unit matrix 1:
\begin{\eq}
\hat S^{\dagger}(\bk) \hat O(\bk) \hat S(\bk) = 1. \label{eq:SOS}
\end{\eq}
We note that matrix $\hat S(\bk)$ cannot be determined uniquely: Using an arbitrary unitary matrix $\hat X$, $\hat S' = \hat S \hat X$ also satisfies eq. (\ref{eq:SOS}).
Here, we introduce the following new basis $(\bar c_{\bk \alpha},\bar c^{\dagger}_{\bk \alpha})$ using $\hat S(\bk)$:
\begin{\eq}
\bar c_{\bk \alpha} = \sum_{\beta} S^{-1}_{\alpha\beta}(\bk) c_{\bk \beta}.
\end{\eq}
We can easily verify that these operators $(\bar c_{\bk \alpha},\bar c^{\dagger}_{\bk \alpha})$ satisfy the canonical anticommutation relations $\{ \bar c_{\bk \alpha}, \bar c^{\dagger}_{\bk' \beta} \} = \delta_{\bk \bk'} \delta_{\alpha\beta}$.
In this basis, eq. (\ref{eq:ham}) is rewritten as 
\begin{\eq}
\hat H_0 = \sum_{\bk,\alpha,\beta} \bar c^{\dagger}_{\bk \alpha} \ldk \bar h_{\alpha\beta}(\bk) -\mu\delta_{\alpha\beta} \rdk \bar c_{\bk \beta}, 
\end{\eq}
where $\bar h_{\alpha\beta}(\bk)= (\hat S^{\dagger}(\bk) \hat h(\bk) \hat S(\bk))^{\alpha\beta} \equiv (\hat {\bar h} (\bk) )^{\alpha\beta}$. 
Therefore, the Green function in the $(\bar c_{\bk \alpha},\bar c^{\dagger}_{\bk \alpha})$ basis is given by
\begin{\eq}
\hat {\bar G}(\bk,\w) = \lk  \w + \mu - \hat {\bar h}(\bk)  \rk^{-1}.  
\label{eq:G-func}
\end{\eq}

Next, we derive the expression for the current operator in the 
$(c_{\bk \alpha}, c^{\dagger}_{\bk \alpha})$ basis. From the continuity equation $\f{\rd}{\rd t} n(\bm{r}) + \nabla \cdot \bm{j} (\bm{r})=0$,
we obtain
\begin{\eq}
\f{\rd}{\rd t} n(\bq) = -i \bq \cdot \bm{j}(\bq),
\end{\eq}
where $n$ is electron number density.
$\f{\rd}{\rd t} n(\bq)$ can be calculated by the equation of motion: $\f{\rd}{\rd t} n(\bq) = i \ldk H,n(\bq) \rdk$.
Therefore, $x$-component of the current operator $j_{x}$  is given by
\begin{\eq}
j_x = \lim_{q_x \rightarrow 0} \lk -\f{1}{q_x} \ldk H,n(\bq) \rdk  \rk.
\label{eq:current}
\end{\eq}
In Appendix \ref{level:App-A}, we show that $n(\bq)$ is given by
\begin{\eq}
n(\bq) = \sum_{\bk,\alpha,\beta} O^{-1}_{\alpha\beta}(\bk) c^{\dagger}_{\bk-{\bq/2}, \alpha}c_{\bk+{\bq/2}, \beta}, \label{eq:nq}
\end{\eq}
which is an exact expression for the first order of $|\bq|$.
Using the following relationship:
$[AB,CD] =A \{B,C\} D - AC\{B,D\} + \{A,C\} DB -C\{A,D\} B,$
$\ldk H,n(\bq) \rdk$ is given by
\begin{align}
\ldk H, n(\bq) \rdk &= \sum_{\bk,\alpha,\beta} c^{\dagger}_{\bk-\bq/2,\alpha} \{ ( \hat h (\bk-\bq/2) \hat O^{-1} (\bk-\bq/2) \hat O(\bk) )^{\alpha\beta} \nn
&- (\hat O(\bk) \hat O^{-1} (\bk+\bq/2) \hat h (\bk+\bq/2)  )^{\alpha\beta} \} c_{\bk+\bq/2, \beta}.
\label{eq:CCR}
\end{align}
By substituting eq. (\ref{eq:CCR}) into eq. (\ref{eq:current}), we obtain the expression for the velocity in the ($c_{\bk \alpha}, c^{\dagger}_{\bk \alpha}$) basis as follows: 
\begin{\eq}
\hat v_{x}(\bk) = \f{\rd \hat h (\bk)}{\rd k_x} + \f{1}{2} \hat h (\bk) \hat D_x(\bk) + \f{1}{2} \hat D^{\dagger}_x(\bk) \hat h (\bk) ,
\label{eq:velocity}
\end{\eq}
where $D_x(\bk)$ is given by
\begin{\eq}
\hat D_x (\bk)&=& \{ \f{\rd}{\rd k_x} \hat O^{-1}(\bk) \} \hat O(\bk) = -\hat O^{-1}(\bk) \f{\rd}{\rd k_x} \hat O(\bk) .  \nn
\end{\eq}
Apparently, $\hat D_x (\bk)$=0 in an orthogonal basis. We call the second and the third terms in eq. (\ref{eq:velocity}) the overlap integral current.
In the ($\bar c_{\bk \alpha}, \bar c^{\dagger}_{\bk \alpha}$) basis, the velocity $\hat {\bar v}_x(\bk)$ is given by 
\begin{\eq}
\hat {\bar v}_x(\bk)= \hat S^{\dagger}(\bk) \hat{\bar v}_x(\bk) \hat S(\bk). \label{eq:bar-velocity}
\end{\eq}
Therefore, even when the overlap integrals between different sites exist, we can calculate the SHC and OHC in the basis ($\bar c_{\bk \alpha}, \bar c^{\dagger}_{\bk \alpha}$) using the matrix $\hat S(\bk)$. In this basis, Green function $\hat {\bar G}(\bk,\w)$ is given by eq. (\ref{eq:G-func}). 
The charge current operator $\hat {\bar J}^C_{\mu}$ and the spin current operator $\hat {\bar J}^S_{\mu}$ are given by 
\begin{\eq}
\hat {\bar J}^C_{\mu} = 
\left(
\begin{array}{cc}
\bar v_{\mu} & 0 \\
0 & \bar v_{\mu} 
\end{array}
\right), \ \ \ \hat {\bar J}^S_{\mu} = 
\left(
\begin{array}{cc}
\bar v_{\mu} & 0 \\
0 & - \bar v_{\mu} 
\end{array}
\right).
\end{\eq}
%Next, the charge current operator $\hat {\bar J}^C_{\mu}$ is given by substituting $\hat {\bar v}_{\mu}(\bk)$ given by eq. (\ref{eq:bar-velocity}) for $\hat v(\bk)$ in eq. (\ref{eq:chargecurrent}). 
Also, the $l_z$-orbital current is given by $\hat {\bar J}^O_{\mu} = \{ \hat {\bar J}^C_{\mu},\hat {\bar l}_z \}/2$ where $\hat {\bar l}_z = \hat S^{\dagger}(\bk) \hat l_z \hat S(\bk)$. Therefore, SHC and OHC can be calculated by substituting $\hat G(\bk,\w), \hat J^C_{\mu}$ and $\hat J^S_{\mu}$ in eqs. (\ref{eq:sxyI}) and (\ref{eq:sxyII}) with $\hat {\bar G}(\bk,\w), \hat {\bar J}^C_{\mu}$ and $\hat {\bar J}^S_{\mu}$, respectively.

%%%%%%%%%%%%%%%%%%%%%%%%%%%%%%%%%%%%%%%%%%%%%%%%
\begin{figure}[!htp]
\includegraphics[width=1.0\linewidth]{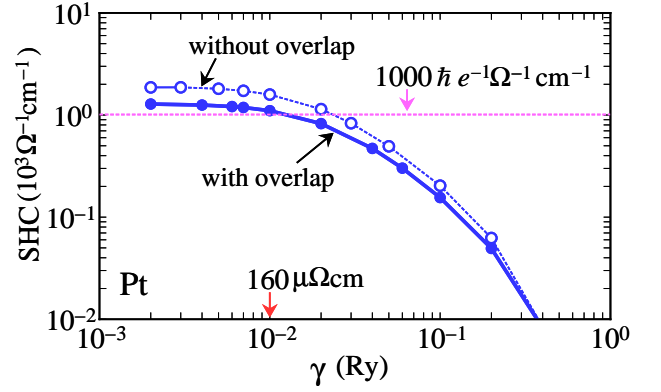}
\caption{\label{fig:compare} 
$\g$-dependence of SHC in Pt obtained by using eq. (\ref{eq:o1}) (with overlap) and that obtained by using eq. (\ref{eq:o0}) (without overlap). When the overlap integrals between different sites are considered, the magnitude of SHC is reduced by about half. The resistivity that corresponds to $\g=0.01$ is $\sim160\mu\Omega$cm in Pt.}
\end{figure}
%%%%%%%%%%%%%%%%%%%%%%%%%%%%%%%%%%%%%%%%%%%%%%%%%%%%

Figure \ref{fig:compare} shows the obtained SHCs in Pt by considering the overlap integrals between different sites in NRL-TB model. For comparison, SHC given by using eq. (\ref{eq:o0}) is also shown.
We find that the magnitude of the SHC is reduced by about half when the overlap integral is considered correctly. We have verified that this is mainly due to the changes of the band spectra, whereas the modification due to the overlap integral current, which is given by the second and third terms in eq. (\ref{eq:velocity}), is less than 10\% in magnitude. Since the CVC is also little affected by the modification of the velocity, we can also safely neglect the CVC when the overlap integrals between different sites are considered.

Here, we comment on the previous study ref. \cite{Kontani-Pt}. Therein, they studied the SHC in Pt using eq. (\ref{eq:o0}). Since the Fermi surface is mainly composed of $d$-electrons in Pt and the $d$-orbital atomic wave functions are well localized, the band structure near the Fermi level is described well in this approximation.
Therefore, the calculations of SHC and OHC in the absence of the overlap integrals between different sites give semiquantitatively reliable results, which can be recognized from Fig. \ref{fig:compare}.
%However, for a more accurate quantitative analyses, we studied the SHE and OHE by taking the overlap integrals between different sites into account correctly. We found that the magnitude of SHC is reduced by about half when these overlap integrals are considered. This is mainly due to the changes of the band spectra, whereas the modification due to the overlap integral current, which is given by the second and third terms in eq. (\ref{eq:velocity}) is less than 10\% in magnitude.

%%%%%%%%%%%%%%%%%%%%%
%Numerical Study
%%%%%%%%%%%%%%%%%%%%%
\section{\label{sec:level4} NUMERICAL STUDY}
\subsection{\label{subsec:level4-1} Numerical results for SHC and OHC}
In this section, we perform the numerical calculations for the SHC and OHC in various 4$d$- and 5$d$- transition metals by considering the overlap integrals between different sites given by eq. (\ref{eq:o1}). In particular, we clarify  SOI ($\ld$), the quasiparticle damping rate ($\g$) and the chemical potential ($\mu$) dependences of the SHC and OHC in each metal. 
%Here, filling $n$ represents the numbers of electron on $4d$-, $5d$-orbitals.
Here, we note that the unit of the SHC and OHC is $|e|/2\pi a$ and 1$[|e|/2\pi a] \approx 1000 \hbar e^{-1} \cdot \Omega^{-1} cm^{-1}$ for $a$=4 $\text{\AA}$.

%First, we shortly comment on the quasiparticle damping rate $\hat \Gamma$ dependence of SHC. 
%In the Born approximation, $\hat \Gamma$ depends on orbital index.
%When $\ld \ll W_{band}$, $\hat \Gamma$ is diagonal with respect to orbital: $\Gamma_{\alpha\beta}=\g_{\alpha} \delta_{\alpha\beta}$, where $\g_{\alpha}$ is proportional to the LDOS for $\alpha$-orbital, $\rho_{\alpha}(0)$. On the other hand, the quasiparticle damping rate $\g_{\alpha}$ is indepedent of orbital in the constant $\g$ approximation: 
%$\Gamma_{\alpha\beta}=\g\delta_{\alpha\beta}$. 
%As reported in ref. \cite{Kontani-Ru}, the intrinsic SHC in Sr$_2$RuO$_4$ is independent of the quasiparticle damping rate in the low resistive regime. However, it depends on the ratio of the quasiparticle damping rate between different orbital states, $\g_{\alpha}/\g_{\beta}$: The SHC in Sr$_2$RuO$_4$ given by the Born approximation is about 3 times larger than that in the constant $\g$ approximation.
%In the present model, we have verified that the SHC in the Born approximation gives the similar results to that in the constant $\g$ approximation in each metal. This fact will be discussed in Appendix \ref{level:App-B}. For this reason, we use the constant $\g$ approximation hereafter. 

%rewritten

%%%%%%%%%%%%%%%%%%%%%%%%%%%%%%%%%%%%%%%%%%
\begin{figure}[!htb]
\includegraphics[width=.8\linewidth]{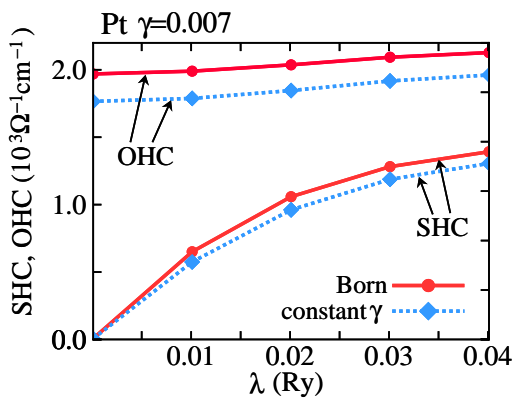}
\caption{\label{fig:born} $\ld$-dependence of SHC and OHC in Pt given by the Born approximation and constant $\g$ approximation. The SHCs obtained in these two approximations give quantitatively similar results. 
We stress that the OHC is finite even if $\lambda=0$.
} 
\end{figure}
%%%%%%%%%%%%%%%%%%%%%%%%%%%%%%%%%%%%%%%%%%
First, we discuss the quasiparticle damping rate $\hat \Gamma$ dependence of SHC.
In the Born approximation, $\hat \Gamma$ depends on orbital index.
When $\ld \ll W_{band}$, $\hat \Gamma$ is diagonal with respect to orbital: $\Gamma_{\alpha\beta}=\g_{\alpha} \delta_{\alpha\beta}$, where $\g_{\alpha}\propto\rho_{\alpha}(0)$.
%is proportional to the LDOS for $\alpha$-orbital, $\rho_{\alpha}(0)$. 
On the other hand, the quasiparticle damping rate $\g_{\alpha}$ is indepedent of orbital in the constant $\g$ approximation: 
$\Gamma_{\alpha\beta}=\g\delta_{\alpha\beta}$.
In Fig. \ref{fig:born}, the SHCs for $\g=0.007$ in these two approximations are shown.
%$\ld$-dependence of the total SHC $\sxy^z=\sxy^{zI} + \sxy^{zII}$ in constant $\g$ approximation with $\g$=0.007 and in the Born approximation. 
We see that the obtained SHC is quantitatively similar in both approximations. 
This fact can be explained as follows:
In transition metals, the LDOS of $t_{2g}$- $(d_{xy},d_{yz},d_{zx})$ and $e_{g}$- $(d_{x^2-y^2},d_{3z^2-r^2})$ orbitals are about equal in magnitude.
Since $\g_{\alpha}\propto\rho_{\alpha}(0)$ in the Born approximation,
two approximation give the similar results in transition metals. 
For this reason, we use the constant $\g$ approximation hereafter.
In contrast, the SHC in $\text{Sr}_2\text{RuO}_4$ given by the Born approximation is much larger than that given by the constant $\g$ approximation since $\alpha$-dependence of $\rho_{\alpha}$ is large \cite{Kontani-Ru}.

%ADD
Figure \ref{fig:born} also shows the OHCs in these two approximations.
As already pointed out in Refs. \cite{Kontani-Ru,Kontani-Pt},
a huge OHC appears even if $\lambda=0$, and it slowly increases
with $\lambda$.
In Sr$_2$RuO$_4$, in contrast, the OHC slowly decreases with $\lambda$ \cite{Kontani-Ru}.
\begin{figure}[!htp]
\includegraphics[width=1.0\linewidth]{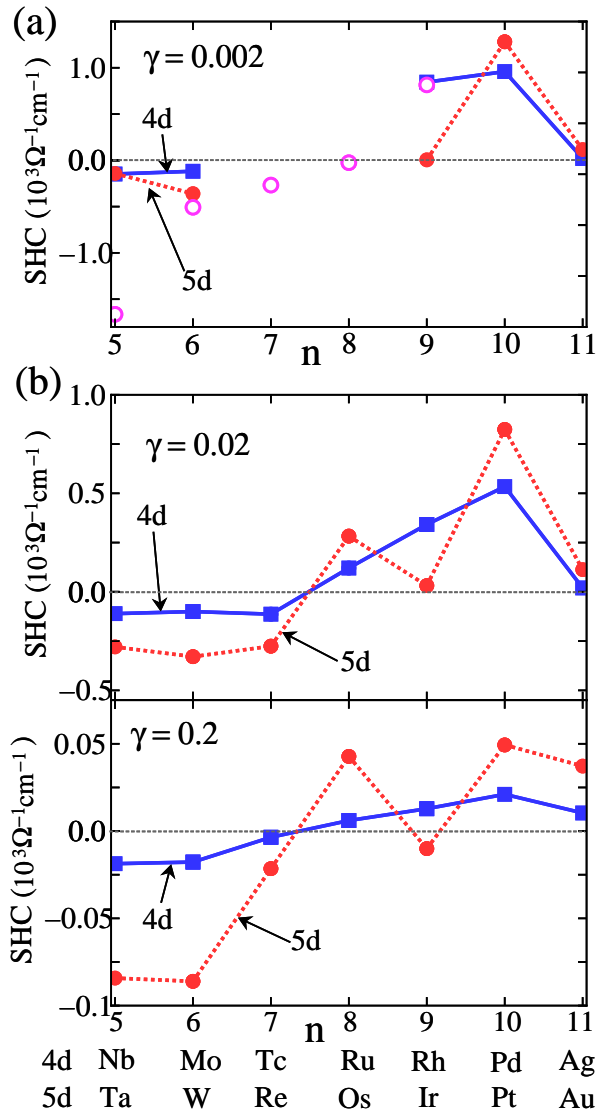}
\caption{\label{n-dep} $n$-dependence of SHC for $\g=0.002$, 0.02 and 0.2. 
In (a), we see that Pt shows the largest SHC in magnitude for $\g=0.002$.
The open symbols represents the SHC in Pt for $n=5\sim 9$.
In (b), the SHCs obtained in the present model for $n=7,8$ (hcp structure) are also shown. 
SHC in W takes the largest value for $\g=0.2$.
}
\end{figure}
%%%%%%%%%%%%%%%%%%%%%%%%%%%%%%%%%%%%%%%%%%

Figure \ref{n-dep} shows the electron number $n$-dependence of the SHC, where $n=n_s + n_d$.
Note that the crystal structure of various transition metals is shown in Table \ref{table1}.
%We note that the crystal structure is given by bcc structure for $n=5,6$, by hcp structure for $n=7,8$, and by fcc structure for $n=9,10,11$. 
The SHC obtained in the present model for $\g=0.002$ is shown in Fig. \ref{n-dep} (a), and for $\g=0.02$ and 0.2 in Fig. \ref{n-dep} (b).
%The SHC in hcp structure for $\g=0.002$ will be studied in a later publication \cite{Kontani-hcp}. 
%bcc structure for  $n=5,6$, hexagonal closed packed (hcp) structure for $n=7,8$, and fcc structure for $n=9,10,11$. The method of calculating SHC in hcp will be explained in the later publication. 
The SHC is negative for $n=5,6$, and positive for $n=9\sim11$: The SHC changes its sign around $n=7$ and 8.
%We see that the sign changes when the filling gets smaller but the absolute value of SHC is quite different from that in Nb, Mo, Ta and W.
The magnitude of SHC is largest in Pt for $\g=0.002$ and 0.02, where the corresponding resistivities are $ \sim 8 \mu\Omega \text{cm}$ and $\sim 64 \mu\Omega \text{cm}$ in Pt, respectively. When $\g=0.2$, however, the absolute values of SHCs in Ta and W become larger than that of Pt, where $\rho\sim 220 \ (250) \ \mu\Omega\text{cm}$ in Pt (Ta and W). Therefore, large negative values of  SHCs in Ta and W will be observed even in high resistive samples.
For comparison, we also calculated the SHC for $n=5 \sim 9$ using the band structure of Pt, which is represented as the open symbols in Fig. \ref{n-dep} (a). 
We see that the magnitude of SHC in Pt with $n=9$ does not reproduce that in Ir. The same is true in Ta and W (bcc structure). Therefore, we need to calculate SHC using a correct band structure for each metal.

Next, we examine the $\ld$-dependence of the SHC. We verified that the SHC in each metal increases approximately proportional to $\ld$ as shown in Fig. \ref{fig:born}. To elucidate the origin of SHC, we calculated the SHC when SOI is anisotropic: $\ld_1\sum (\hat l_z \hat s_z) + \ld_2\sum (\hat l_x \hat s_x + \hat l_y \hat s_y)$. We find that the SHC for $H_{SO}= \ld \sum (\hat l_z \hat s_z)$ $(\ld_1=\ld,\ld_2=0)$ is as large as the SHC in the isotropic case ($\ld_1=\ld_2=\ld$). In contrast, the SHC for $H_{SO}=\ld \sum (\hat l_x \hat s_x + \hat l_y\hat s_y)$ $(\ld_1=0,\ld_2=\ld)$ is one order of magnitude smaller than the isotropic case. Therefore, the $z$-component of the SOI gives the decisive contribution to the SHC.
The matrix element of $\hat l_z$ is finite only for $\langle yz | l_z| zx \rangle= -\langle zx|l_z| yz\rangle = i$ and $\langle xy |l_z | x^2-y^2 \rangle = -\langle x^2-y^2 |l_z | xy \rangle =2i $. 
Note that $d_{xy}$- and $d_{x^2-y^2}$-orbitals ($d_{yz}$- and $d_{zx}$-orbitals) are given by the linear combination of $l_z=\pm 2 \ (l_z=\pm 1)$.
Here, we examine which orbitals among them cause the significant contribution to the SHC.  $z$-component of SOI is rewritten as 
$\displaystyle \ld_3 \sum_{i} \ltk P(l_z^2=1)\left( \hat l_z \hat s_z \right) \rtk_{i} + \ld_4 \sum_{i} \ltk P(l_z^2=4)\left( \hat l_z \hat s_z \right) \rtk_{i}$, where $P(l_z^2=n)$ represents the projection operator. SHC caused by $d_{xy}$- and $d_{x^2-y^2}$-orbitals is given by setting $\ld_3 =0, \ \ld_4=\ld$, which is represented as $l_z=\pm 2$ in Table \ref{table2}. Similarly, SHC caused $d_{yz}$- and $d_{zx}$-orbitals is given by setting $\ld_3 =\ld, \ \ld_4=0$, which is represented as $l_z=\pm 1$ in Table \ref{table2}.
%Table \ref{table2} shows the SHC that arises from these matrix elements. In this table, $d_{xy}$ $\leftrightarrow$ $d_{x^2-y^2}$ represents the SHC caused by $d_{xy}$- and $d_{x^2-y^2}$- orbitals, both of which are given by the linear combination of $l_z=\pm2$. Similarly, $d_{xz}$ $\leftrightarrow$ $d_{yz}$ represents the SHC caused by $d_{zx}$- and $d_{yz}$-orbitals, both of which are given by the linear combination of $l_z=\pm1$.
We see that the interorbital transition between $d_{xy}$- and $d_{x^2-y^2}$- orbitals causes the significant contribution to the SHC in many metals. 
Only in the case of Mo, W and Ir, the contribution of $d_{zx}$- and $d_{yz}$-orbitals is comparable to that of $d_{xy}$- and $d_{x^2-y^2}$-orbitals. In other metals, $d_{xy}$- and $d_{x^2-y^2}$- orbitals give the dominant contribution to the SHC.

%%%%%%%%%%%%%%%%%%%%%%%%%%%%%%%%%%%%%%%%%%
\begin{table}[!htb]
\caption{\label{table2} SHC which originates from the $d_{zx}$, $d_{yz}$, $d_{xy}$ and $d_{x^2-y^2}$-orbitals. Here, we set $\g=0.02$. $l_z=\pm 2$ ($l_z=\pm 1$) represents the SHC caused by $d_{xy}$- and $d_{x^2-y^2}$- orbitals ($d_{yz}$- and $d_{zx}$- orbitals). The ratio represents (SHC from $l_z=\pm 2$) / (SHC from $l_z=\pm 1$). We see that $d_{xy}$- and $d_{x^2-y^2}$- orbitals cause the significant contribution to the SHC in many metals.}
\begin{ruledtabular}
\begin{tabular}{lrrr}
metals & $l_z=\pm 1$ & $l_z=\pm 2$ & ratio\\ \hline
Nb(4$d^4$5$s^1$)& -0.0332 & -0.0770 & 2.32 \\
Mo(4$d^5$5$s^1$) & -0.0474 & -0.0587 & 1.24 \\
Rh(4$d^8$5$s^1$) & 0.0847 & 0.269 & 3.18 \\
Pd( 4$d^{10}$5$s^0$) & 0.0847 & 0.455 & 5.37 \\
Ag(4$d^{10}$5$s^1$) & 0.00224 & 0.0181 & 8.08  \\ \hline
Ta(5$d^3$6$s^2$)  & -0.0222 & -0.254 & 11.4 \\
W(5$d^4$6$s^2$)  & -0.174  & -0.205 & 1.18 \\
Ir(5$d^9$6$s^0$) & 0.0123 & 0.0231 & 1.89 \\
Pt(5$d^{9}$6$s^1$)& 0.136 & 0.678 & 4.98 \\
Au(5$d^{10}$6$s^1$) & 0.0177 & 0.0987 & 5.59 \\
\end{tabular}
\end{ruledtabular}
\end{table}
%%%%%%%%%%%%%%%%%%%%%%%%%%%%%%%%%%%%%%%%%

%%%%%%%%%%%%%%%%%%%%%%%%%%%%%%%%%%%%%%%%%%
\begin{figure}[!htp]
\includegraphics[width=.99\linewidth]{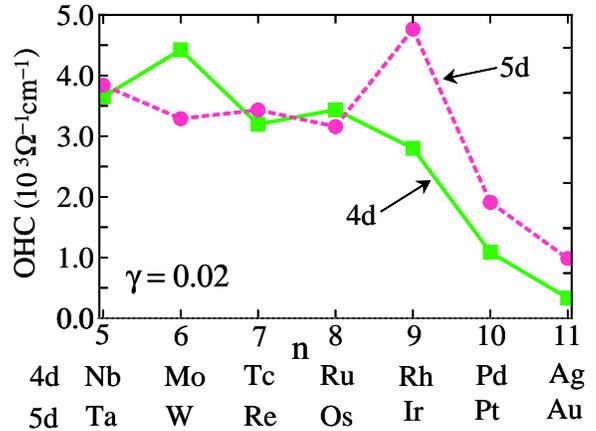}
\caption{\label{fig:OHE} $n$-dependence of OHC for $\g=0.02$. 
The obtained OHCs are positive for all metals, and they are
about 10 times larger than the SHCs except for Pt and Pd.
}
\end{figure}
%%%%%%%%%%%%%%%%%%%%%%%%%%%%%%%%%%%%%%%%%%%

%ADD
Here, we show the OHCs for $\g=0.02$ in various transition metals 
in Fig. \ref{fig:OHE}.
We see that all the $4d$ and $5d$ transition metals show 
huge and positive OHCs,
which is almost one order of magnitude larger than that of the SHCs.
In Au (Ag), the OHC takes a small value since the $d$-electron DOS 
is small at the Fermi level.
Therefore, huge and positive OHC is a universal nature of transition metals.
As in the case of the SHE, the intrinsic OHE shows the crossover behavior:
the OHC independent of $\rho$ in the low resistive regime,
whereas it decreases in proportion to $\rho^{-2}$ in the high
resistive regime \cite{Kontani-Ru,Kontani-Pt}.
In a later publication, we will present an intuitive (semiclassical) 
explanation for the origin of the OHC \cite{future}.
%The OHC is positive in every 4$d$- and 5$d$- transition metals, whereas the sign of SHC depends on the electron number $n$:
%In \ref{subsec:level5-2}, we explain why the OHE occurs independently 
%of the SOI, and the sign of OHC is positive in each transition metal. 
%We discuss that the OHE is the essential phenomenon that is a natural 
%consequence of the asymmetric $s$-$d$ hybridization,
%and the SHE is a passive phenomenon induced by the OHE if SOI presents.
%In this sence, the OHE is the origin of the SHE in transition metals.

%%%%%%%%%%%%%%%%%%%%%%%%%%%%%%%%%%%%%%%%%%
\begin{figure}[!htp]
\includegraphics[width=0.9\linewidth]{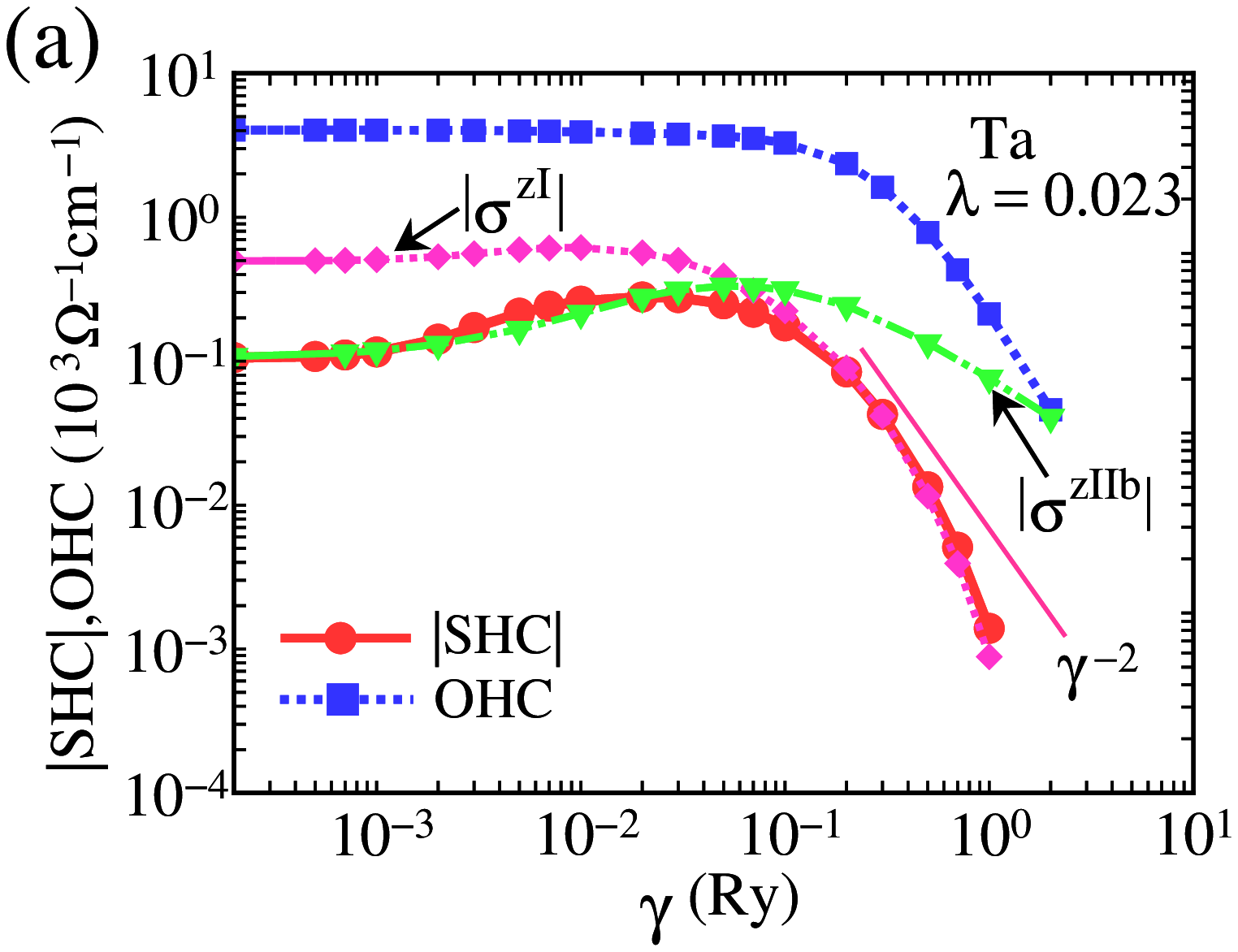}
\includegraphics[width=0.9\linewidth]{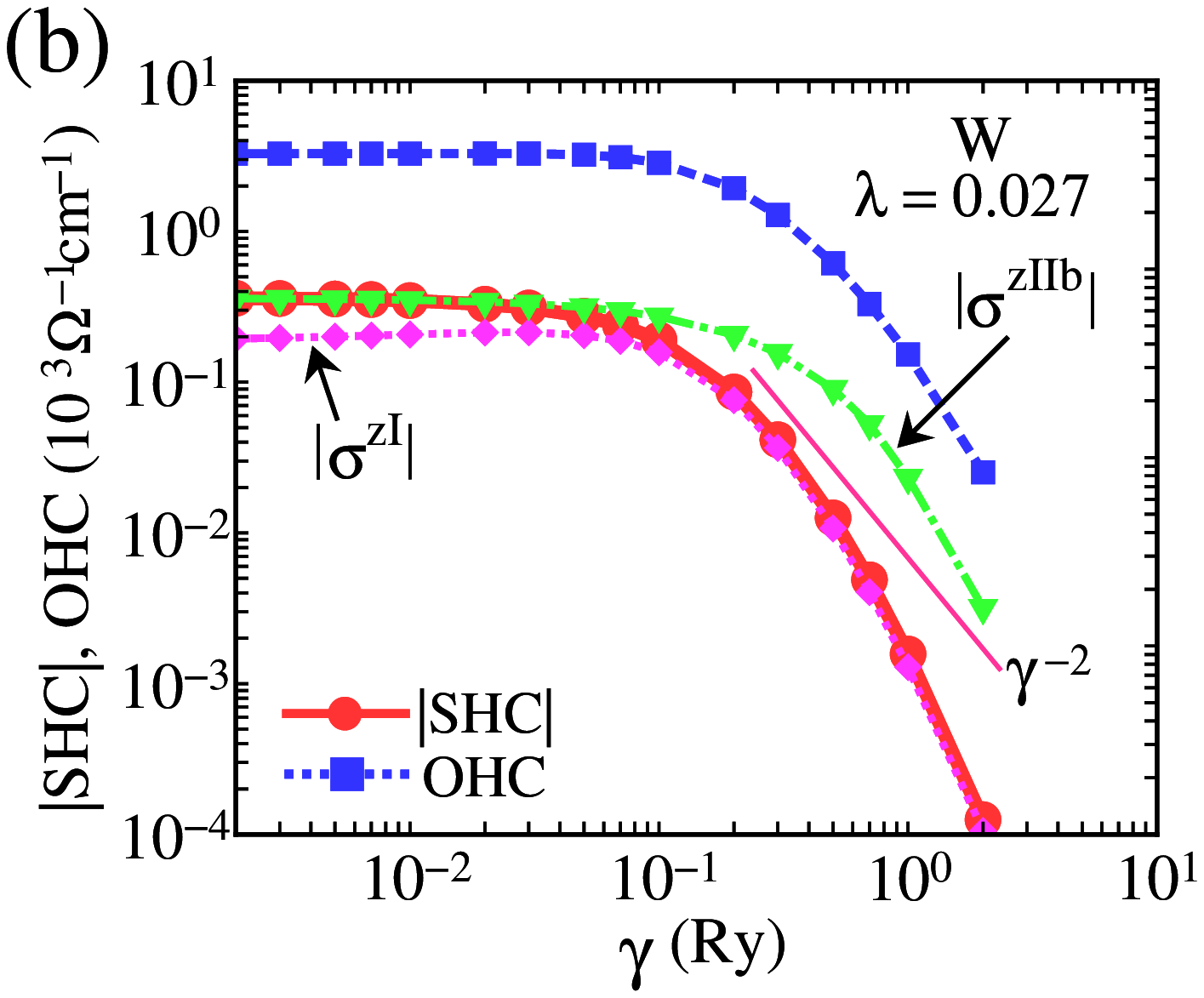}
\caption{\label{fig:Ta} $\gamma$-depnedence of SHC and OHC in Ta (a) and W (b). The corresponding resistivity $\rho$ to $\g=0.1$ is $\sim 190 \ \mu\Omega\text{cm}$ in Ta and $\sim 220 \ \mu\Omega\text{cm}$ in W. }
\end{figure}
%\begin{figure}[!htp]
%\includegraphics[width=.85\linewidth]{W-gamma.eps}
%\caption{\label{fig:W} $\gamma$-depnedence of SHC and OHC in W.}
%\end{figure}
%%%%%%%%%%%%%%%%%%%%%%%%%%%%%%%%%%%%%%%%%%
Now, we discuss the $\g$-dependences of SHC and OHC. The $\g$-dependence of intrinsic SHCs in Ta and W are shown in Fig. \ref{fig:Ta}. 
In usual, the intrinsic Hall conductivities are independent of $\g$ in the low resistive regime where $\g\ll \Delta$, whereas it decreases approximately in proportion to $\rho^{-2}$ in the high resistive regime where $\Delta\ll \g$ \cite{{Kontani94},{Kontani06}}: Here, $\Delta$ represents the band-splitting measured from the Fermi level. In W, $\Delta$ is $\sim 0.04$, and in Pt, $\Delta$ is $\sim 0.035$ \cite{Kontani-Pt}.
We find that W shows a typical coherent-incoherent crossover at $\g\sim \Delta$ in Fig. \ref{fig:Ta} (b). 
%The coherent-incoherent crossover is also realized in Pt, as shown in Fig. \ref{fig:compare} (a).
We have also verified that the coherent-incoherent crossover behavior is universally seen in many transition metals including Pt, which is shown in Fig. \ref{fig:compare} (a).
However, as shown in Fig. \ref{fig:Ta} (a), SHC in Ta shows an exceptional behavior: It takes a maximum value around $\gamma\sim 0.02$ and decreases as $\g$ decreases in the low resistive regime. We find that this anomalous behavior can arise when almost degenerate anticrossing points exist slightly away from the Fermi level. We will discuss the reason in detail in section \ref{subsec:level4-2}.

In fig. \ref{fig:Ta} (b), the $\g$-dependences of the Fermi surface term $\sxy^{zI}$ and the Fermi sea terms $\sxy^{zIIa}$,$\sxy^{zIIb}$ are also shown. In the low resistive regime, the relation $\sxy^{zI}\simeq \sxy^{zIIb}$ holds well, and $\sxy^{zIIb}$ reproduces the total Hall conductivity $\sxy^z$ \cite{Murakami-SHE}. In the high resistive regime, however, $\sxy^z\simeq\sxy^{zI}$ whereas $\sxy^{z}$ is quite different from $\sxy^{zIIb}$ in the high resistive regime. As a result, the relationship 
\begin{\eq}
\sxy^z\simeq\sxy^{zI} \ \ \text{(Fermi surface term)}, \label{eq:FS}
\end{\eq}
is recognized for a wide range of $\g$. We will discuss the crossover behavior of the intrinsic Hall conductivity in more detail in section \ref{subsec:level5-3}.

We also discuss the $\g$-dependence of OHC. The obtained OHCs in Ta and W are shown in Fig. \ref{fig:Ta}.
The coherent-incoherent crossover behavior of OHC is recognized in Fig. \ref{fig:Ta}. In contrast to the $\g$-dependence of SHC in Ta, OHC shows a typical crossover behavior. We also verified that OHC is finite even if $\ld=0$ since the $d$-orbital current in eq. (\ref{eq:Jo}) is independent of the spin index \cite{Kontani-Ru}.

\begin{figure}[!htp]
\includegraphics[width=0.75\linewidth]{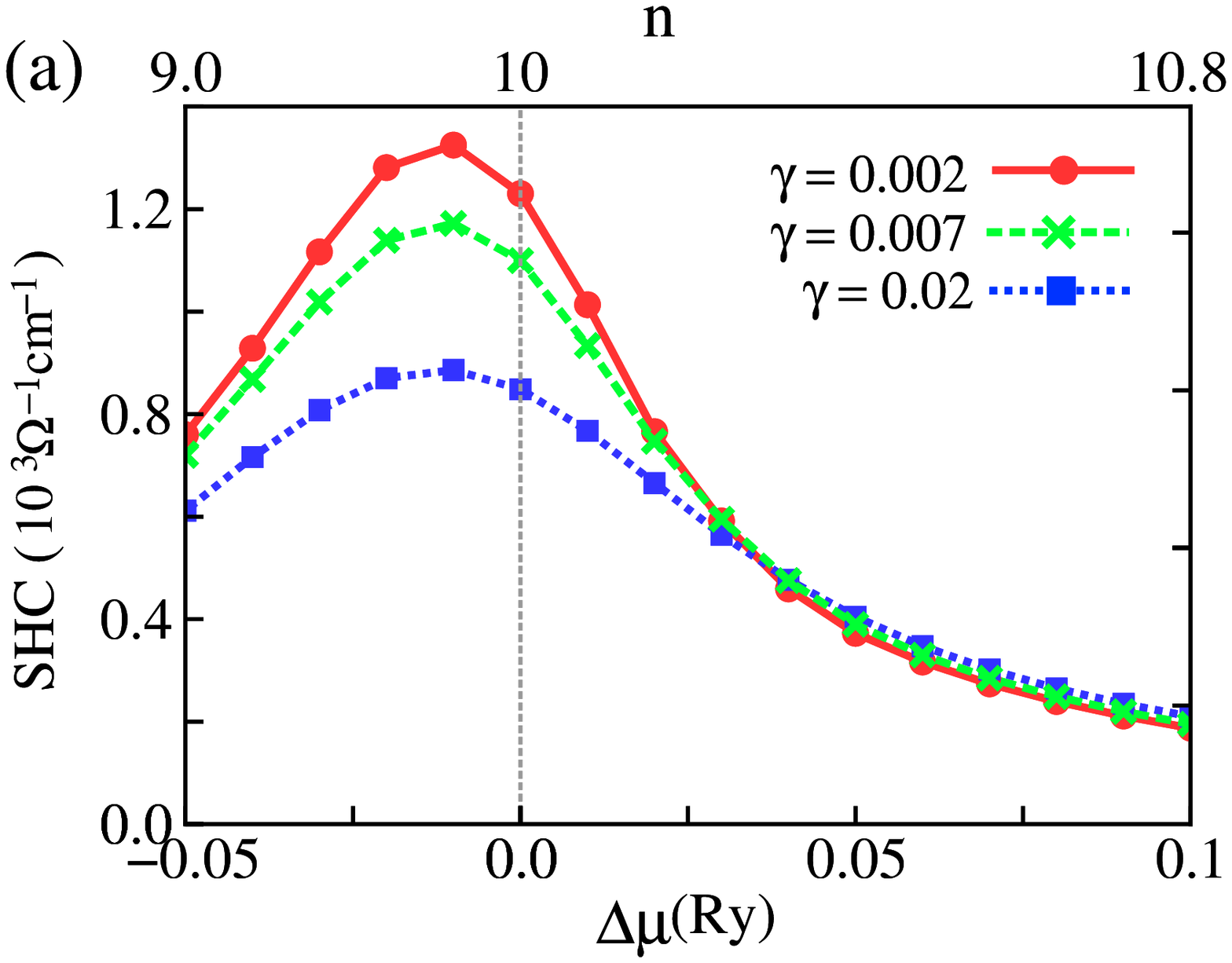}
\includegraphics[width=0.75\linewidth]{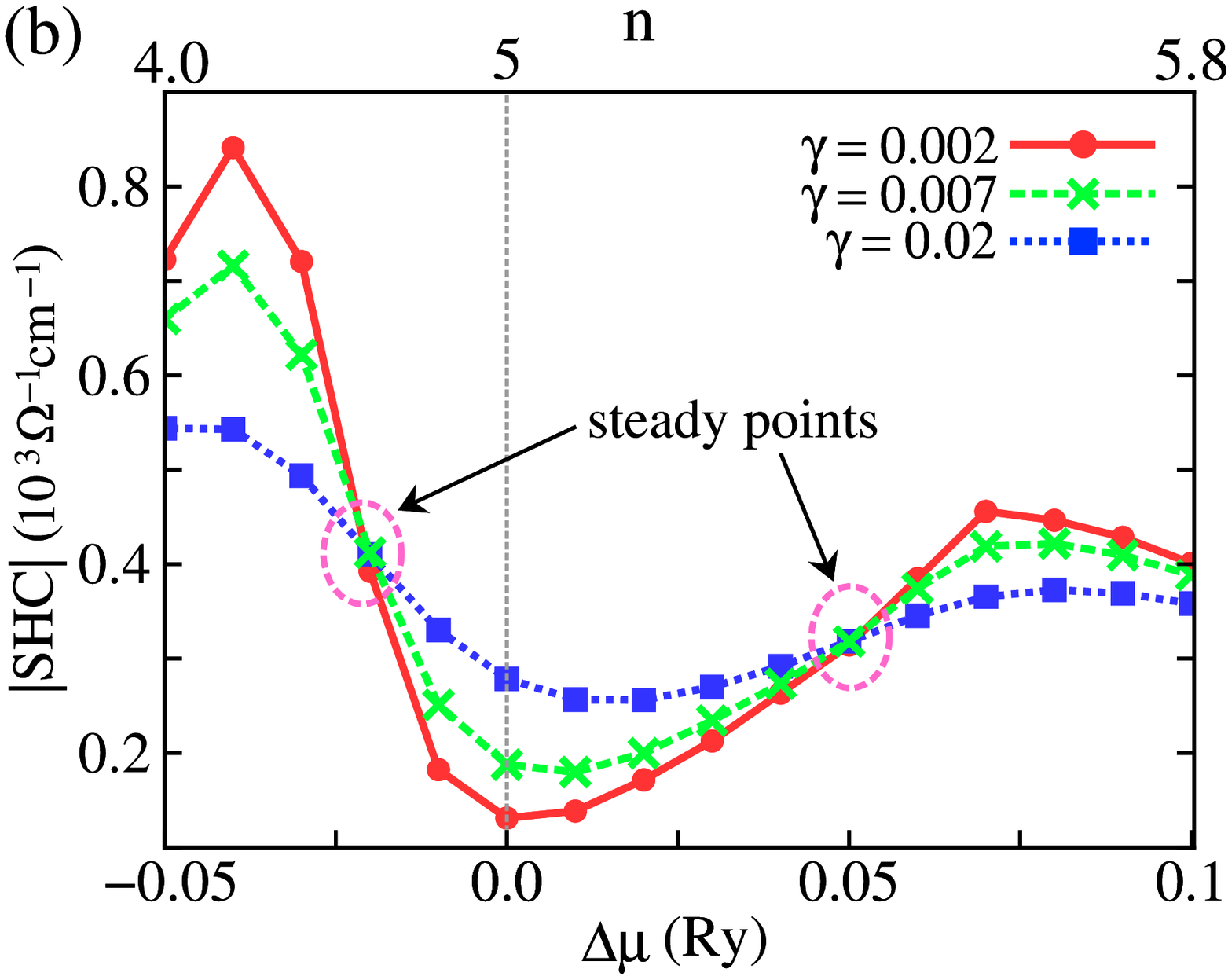}
\caption{\label{fig:mu-dep-Pt} 
$\mu$-dependence of $\sxy^{zIIb}$ in (a) Pt and (b) Ta for $\g=0.002, 0.007,0.02$. The chemical potential is given by $\mu$=$\mu_0$+$\dm$ where $\mu_0$ represents the true value of chemical potential. 
Total electron number $n$ is shown in the upper horizontal axis.
%In Pt, the changes of Fermi eneregy $\Delta\mu=-0.047, 0$, and $0.17$ corresponds to the electron number $n=9,10$, and 11, respectively.
Note that the sign of the SHC in Ta is negative.
}
\end{figure}

Now, we discuss the $\mu$-dependence of SHC in Pt and Ta assuming that the bandstructure is rigid.
Experimentally, the chemical potential $\mu$ can be controlled by composing alloys.
Figure \ref{fig:mu-dep-Pt} (a) shows the $\mu$-dependence of $\sxy^{zIIb}$ in Pt for $\g=0.002, 0.007$ and 0.02. 
Here, the chemical potential is given by $\mu$= $\mu_0$+ $\dm$, where $\mu_0$ represents the true value of chemical potential.
%add
%In Pt, changing the Fermi energy as $\Delta\mu=-0.11, -0.047, 0$, and $0.17$ corresponds to change the electron numbers as $n=8,9,10$, and 11, respectively.
%
We see that the SHC shows a peak around $\dm=0$ and it decreases when $\mu$ is raised or lowered from its true value. 
%The SHC becomes negative for $\Delta\mu < -0.11$.  
This $\dm$-dependence of SHC obtained in the present model in Pt seems to be in good agreement with that in ref. \cite{Guo-Pt}. We see that SHC for $\g=0.002$ is about 45$\%$ larger than that for $\g$=0.02 at $\Delta\mu$=0. 
%However, as $|\dm|$ increases, the magnitude of SHCs for $\g=0.002$ and 0.02 begin to approach each other, and become the same around $\dm=-0.09$ and $\dm=+0.04$. 
%We find that several ``steady points", where the magnitude of SHC is approximately independent of $\g$ exist in each metal in the low resistive regime.

%%%%%%%%%%%%%%%%%%%%%%%%%%%%%%%%%%%%%%%%
%%%%%%%%%%%%%%%%%%%%%%%%%%%%%%%%%%%%%%%%

%%%%%%%%%%%%%%%%%%%%%%%%%%%%%%%%%%%%%%%%%%%
\begin{figure}[!htp]
\includegraphics[width=.85\linewidth,height=.5\linewidth]{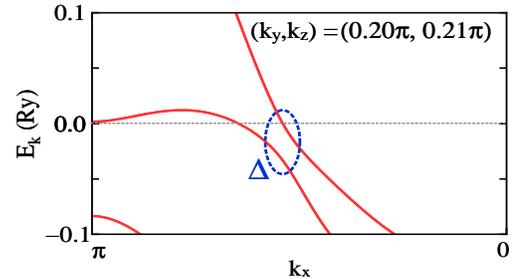}
\caption{\label{fig:Pt-disp} The $k_x$-dependence of $E^{l}_{\bk}$ for $(k_y,k_z)=(0.20\pi,0.21\pi)$ in Pt is shown. The bandsplitting measured from the Fermi level is $\Delta\sim 0.035$. 
A wide area around $(k_x,k_y,k_z)$=$(0.54\pi,0.20\pi,0.21\pi)$ gives a dominant contribution to SHC in Pt.}
\end{figure}
%%%%%%%%%%%%%%%%%%%%%%%%%%%%%%%%%%%%%%%%%

Here, we elucidate from which part of the surface the SHC in Pt originates by calculating $\displaystyle \sxy^{z}(\bk)\equiv \f{1}{8} \sum^{(\pm k_x,\pm k_y,\pm k_z)}_{k^{'}_x,k^{'}_y,k^{'}_z} \text{Tr} \ldk \hat J^S_x \hat G^R \hat J^C_y \hat G^A \rdk_{\bk',\w=0} $. Note that $\f{1}{2\pi N}\sum_{\bk}\sxy^z(\bk)$ is equal to eq. (\ref{eq:sxyI}). 
$\sxy^{z}(\bk)$ is finite only on the Fermi surface, and it takes a huge value at (0.76$\pi$,0,0) (on $\Gamma$-X) and at (0.44$\pi$,0.44$\pi$,0.44$\pi$) (on L-$\Gamma$) since two bands are very close near the Fermi level in the present model \cite{Kontani-Pt}: $\sxy^z(\bk)\sim 3000$ for the former point and $\sxy^z(\bk)\sim 5000$ for the latter point in the present model. However, the contribution of these points to the SHC is small after taking $\bk$-summation using $512^3$ $\bk$-meshes.
We verified that the dominant contribution comes from a wide area around $(k_x,k_y,k_z)$=$(0.54\pi,0.20\pi,0.21\pi)$ as shown in Fig. \ref{fig:Pt-disp}.

Figure \ref{fig:mu-dep-Pt} (b) shows the $\mu$-dependence of $\sxy^{zIIb}$ in Ta for $\g=0.002, 0.007$ and 0.02. 
%add
%In Ta, changing the Fermi energy as $\Delta\mu=-0.05, 0$, and $0.13$ corresponds to change the electron numbers as $n=4,5$, and 6, respectively.
%
%SHC in Ta remains negative in the regime $\dm\in$ [-0.08, 0.1]. 
%We see that the magnitude of SHC for $\g$=0.02 is larger than that for $\g$=0.002 at $\mu$=0. However, when $\mu$ is varied from $\mu=0$, the manitude of SHCs for these two $\g$ exchange across the ``steady points''.
In Ta, several ``steady points", where the magnitude of SHC is approximately independent of $\g$ in the low resistive regime, are recognized at $\dm=-0.02$ and $\dm=+0.05$ in Fig. \ref{fig:mu-dep-Pt} (b). 
When $\mu$ decreases across the steady point at $\dm=-0.02$, the magnitude of SHC increases and reaches a peak around $\dm=-0.04$. This peak originates from the almost degenerate anticrossing bands, which is discussed in more detail in section \ref{subsec:level4-2}.
%We find that this peak originates from the Dirac cone type dispersion which is shown in Fig. \ref{fig:Ta-dc-disp}. When $\mu$ is around $\mu-0.04$, the Fermi level begins to lie inside the Dirac cone type dispersion. 

%%%%%%%%%%%%%%%%%%%%%%%%%%%%%%%%%%%%%%%%%%
%\begin{figure}[!htp]
%\includegraphics[width=1.0\linewidth]{Fig10.eps}
%\includegraphics[width=.55\linewidth]{Fig10.eps}
%\caption{\label{fig:mu-dep-Ta} $\mu$-dependence of $\sxy^{zIIb}$ in Ta for $\g=0.002, 0.007,0.02$. There is a sharp  peak around $\dm=-0.04$. 
%In Ta, the changes of Fermi eneregy $\Delta\mu=-0.05, 0$, and $0.13$ corresponds to the electron number $n=4,5$, and 6, respectively. 
%Note that the sign of the SHC in this figure is negative.
%}
%\end{figure}
%%%%%%%%%%%%%%%%%%%%%%%%%%%%%%%%%%%%%%%%%%

Finally, we explain why SHC in Ir is small in magnitude, by analysing $\sum_{\bk}\sxy^{z}(\bk)$. Here, we divide the $\bk$-summation into $\bk^{+}$-region and $\bk^{-}$-region, where $\bk^+$-region ($\bk^-$-region) represents the region where $\sxy^z(\bk)>0$ ($\sxy^z(\bk)<0$) holds: 
\begin{align}
\sum_{\bk} \sxy^{z}(\bk) &= \sum_{\bk^+} \sxy^z(\bk) + \sum_{\bk^-}\sxy^z(\bk) \nn
                                 &\equiv \sxy^{z+} + \sxy^{z-}.
\end{align} 
In many transition metals, such as Pt and Ta, either $\sxy^{z+}$ or $|\sxy^{z-}|$ is much larger than the other.
In Ir, however, we have verified that the relation $\sxy^{z+}\simeq  |\sxy^{z-}|$ holds, and therefore $\sxy^z$ becomes small in magnitude.

%%%%%%%%%%%%%%%%%%%%%%%%%%
%subsection 4-2
%%%%%%%%%%%%%%%%%%%%%%%%%%
\subsection{\label{subsec:level4-2} Mechanism of impurity assisted SHC}
In the previous section, we have verified that the SHC in all 4$d$- and 5$d$- transition metals except for Ta are independent of $\g$ in the low resistive regime ($\g\ll\Delta$). 
Here, we show that the SHC can show nonmonotonic $\g$ dependence in the low resistive regime
when almost degenerate anticrossing points exist slightly away from the Fermi level,
due to the impurity-assisted interband excitation.
This is the origin of the anomalous $\g$-dependence of the SHC in Ta for $\g < 0.02$ in Fig. \ref{fig:Ta} (a). We call this phenomenon the impurity-assisted SHE.

\begin{figure}[!htp]
\includegraphics[width=.8\linewidth]{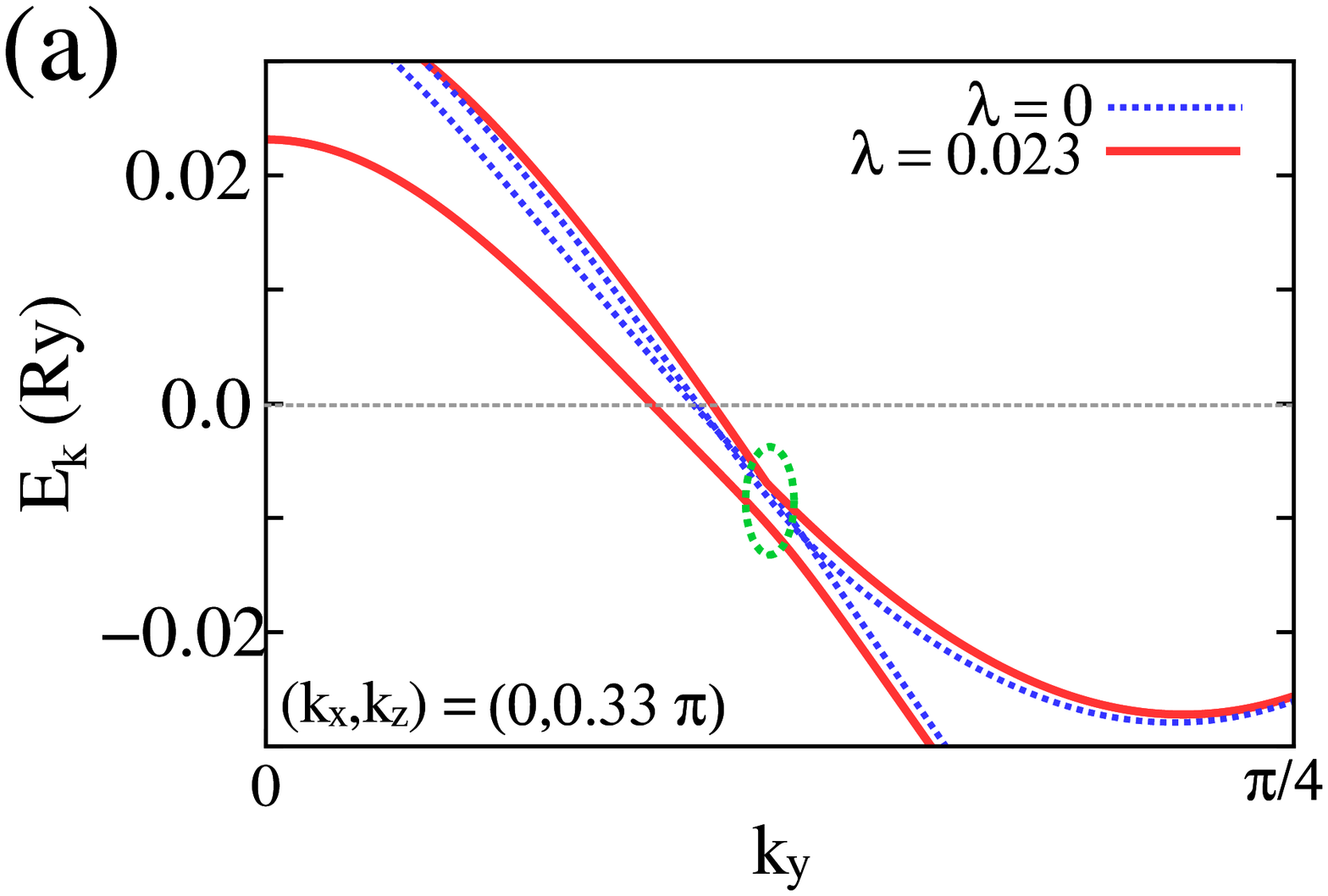}
\includegraphics[width=.6\linewidth]{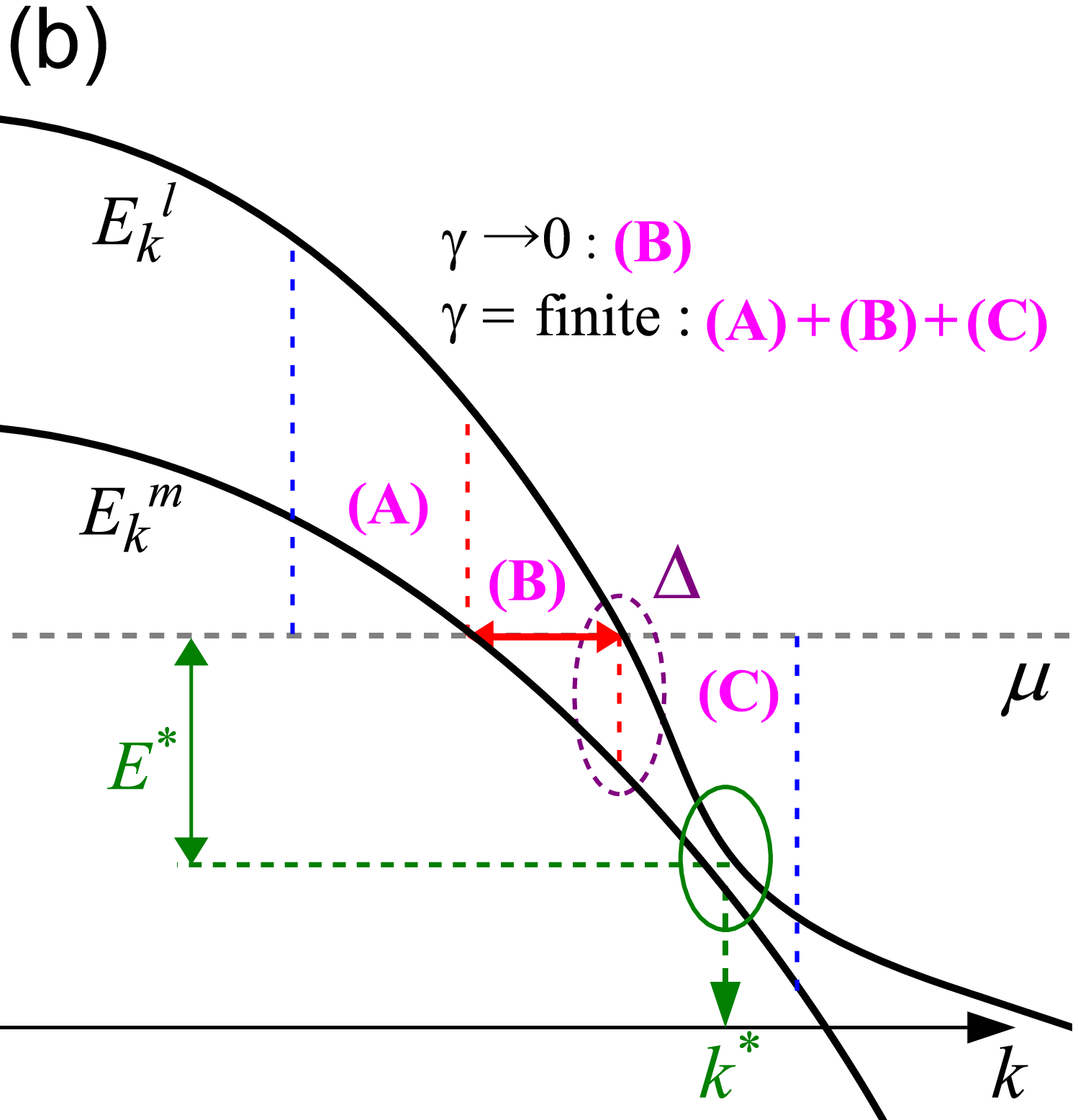}
\caption{\label{fig:IIb-term} 
(a) Anticrossing bands in NRL-TB model for Ta. We find an accidental degenerate point at $(k_x,k_y,k_z)$=$(0,0.12\pi,0.33\pi)$ in the present model with $\ld$=0. The bandsplitting at $\bk^{\ast}$ is $\sim$ 0.003.
(b) Band structure around an accidental degenerate point slightly away from the Fermi level. The region where $\bk$-summation is performed for $\g\rightarrow 0$ and finite $\g$ are represented by using (A), (B) and (C). $\Delta$ represents the band-splitting measured from the Fermi level and $\bk^{\ast}$ represents the point of the minimum band-splitting around the accidental degenerate point.
}
\end{figure}

In Ta, there are several accidental degenerate points with $\ld$=0 slightly away from the Fermi level.
We show the anticrossing bands of NRL-TB model for Ta in Fig. \ref{fig:IIb-term} (a). We find an accidental degenerate point at $(k_x,k_y,k_z)$=$(0,0.12\pi,0.33\pi)$ in the present model with $\ld$=0. Note that this degeneracy splits with $\ld\neq 0$, as recognized in Fig. \ref{fig:IIb-term} (a).

%%%%%%%%%%%%%%%%%%%%%%%%%%%%%%%%%%%%%%%%%%
%%%%%%%%%%%%%%%%%%%%%%%%%%%%%%%%%%%%%%%%%

In usual band structures, we have shown that dominant contribution arises from the Fermi surface term $\sxy^{zI}$ and the relation given by eq. (\ref{eq:FS}) holds well. On the other hand, in the exceptional case such as in Ta, the anomalous $\g$-dependence of SHC can be explained by analyzing $\sxy^{zIIb}$ as follows:
By dropping the current operators in eq. (\ref{eq:IIb}) for simplicity, $\sxy^{zIIb}$ is given by
\begin{\eq}
\sxy^{zIIb}\propto \sum_{\bk, l > m} \f{1}{(E^l_{\bk}-E^{m}_{\bk})^2} \text{Im}\ltk \ln \lk \f{E^l_{\bk}-i\g}{E^m_{\bk}-i\g} \rk \rtk. \label{eq:IIb-approx0}
\end{\eq}
We can approximate as follows for $\g\ll |E^l_{\bk}|, |E^m_{\bk}|$:
\begin{align}
&\text{Im} \ltk \ln \lk \f{E^l_{\bk}-i\g}{E^m_{\bk}-i\g} \rk \rtk \approx -\pi\theta(-E^l_{\bk})+ \pi\theta(-E^m_{\bk}) \nn
& \qquad \qquad \qquad + \f{\g (E^l_{\bk}-E^m_{\bk})}{E^l_{\bk}E^m_{\bk}}.
\end{align}
Substituting above equation into eq. (\ref{eq:IIb-approx0}), we obtain the following relation for small $\g$:
\begin{align}
\sxy^{zIIb}\propto \sum_{\bk, l > m} &\ldk \f{ \theta (-E^m_{\bk} ) -\theta ( -E^l_{\bk}) }{ (E^l_{\bk}- E^m_{\bk})^2 } \right. \nn
&\left. + \g\f{\ \theta(|E^{l}_{\bk}|-\g)\theta(|E^{m}_{\bk}|-\g)}{E^l_{\bk}E^m_{\bk}(E^l_{\bk}-E^m_{\bk})}  \rdk, \label{eq:IIb-approx1}
\end{align}
where the step function of the second term in above equation is introduced to skip the $\bk$-summation in the case of $\g\geq  |E^l_{\bk}|, |E^m_{\bk}|$.

Figure \ref{fig:IIb-term} (b) is a schematic band structure around the accidental degenerate point slightly away from the Fermi level. In this figure, $\Delta$ represents the band-splitting measured from the Fermi level and $E^{\ast}$ represents the eigenenergy measured from the Fermi level: $E^{\ast}=E^{l}_{\bk^{\ast}}\simeq E^{m}_{\bk^{\ast}}$.

%Using Fig. \ref{fig:IIb-term} (b), we discuss the $\g$-dependence of eq. (\ref{eq:IIb-approx1}). 
The first term of eq. (\ref{eq:IIb-approx1}) is finite only in region (B) where $E^l_{\bk}>0$ and $E^m_{\bk}<0$, and its sign is positive
%However, it vanishes in regions (A) and (C) where $E^l_{\bk}\cdot E^m_{\bk} <0$. The sign of the first term in region (B) is positive.
The sign of the second term is negative in region (B), whereas it is positive in regions (A) and (C).
When $\g\rightarrow 0$, $\sxy^{zIIb}$ is given only by the first term in eq. 
(\ref{eq:IIb-approx1}) since the second term vanishes. 
%The second term in eq. (\ref{eq:IIb-approx1}) contribute to $\sxy^{zIIb}$ only when $\g$ is finite. 
When $E^{\ast}\sim \g$, a large contribution to $\sxy^{zIIb}$ comes from the second term in eq. (\ref{eq:IIb-approx1}) from the regime (C) in Fig. \ref{fig:IIb-term} (b). 
Since the second term can be as large as the first term, $\sxy^{zIIb}$ takes the sizable peak at finite $\g$ 
%in Ta. 
%Therefore, the second term in eq. (\ref{eq:IIb-approx1}) gives rise to the anomalous $\g$-dependence of SHC 
in the presence of almost degenerate anticrossing points near the Fermi level. The SHC reaches the maximum value at $\g\sim E^{\ast}$.
As a result, the anomalous behavior of SHC in NRL-TB model for Ta originates from the anticrossing points as shown in Fig. \ref{fig:IIb-term} (a). From this figure, we see that $E^l_{\bk^{\ast}}$ is $\sim 0.01$. This fact is consistent with the peak of SHC $\sxy^z$ around $\g=0.02$ in Fig. \ref{fig:Ta} (a).
%dominant contribution?

\begin{figure}[!htp]
\includegraphics[width=.8\linewidth]{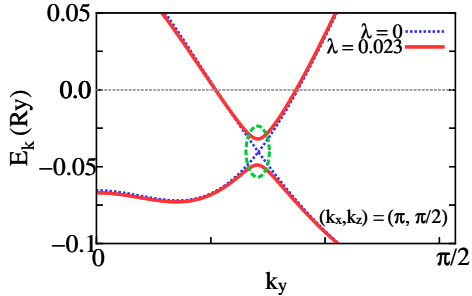}
\caption{\label{fig:Ta-dc-disp} Anticrossing bands near the Fermi level with $\ld=0,0.023$ in Ta: The $k_y$-dependence of $E^{l}_{\bk}$ for $(k_x,k_z)=(\pi,\pi/2)$ is shown. In the present model with $\ld=0$, we find an accidental degenerate point at $(k_x,k_y,k_z)$=$(\pi,0.23\pi,\pi/2)$, which is different point from that shown in Fig. \ref{fig:IIb-term} (b).
The corresponding minimum bandsplitting induced by SOI is $\sim0.015$. }
\end{figure}

In the present model with $\ld$=0, there is another accidental degenerate point at $(k_x,k_y,k_z)$=$(\pi,0.23\pi,\pi/2)$. The band structure obtained for NRL-TB model in Ta with $\ld$=0 and $\ld$=0.023 around this point is shown in Fig. \ref{fig:Ta-dc-disp}.
From this figure, when $\mu$ is lowered to $\dm=-$0.04, we see that the Fermi level begins to lie inside the gap. This fact causes a sharp peak of $\sxy^{zIIb}$ at $\dm=-$0.04 in Fig. \ref{fig:mu-dep-Pt} (b).

%%%%%%%%%%%%%%%%%%%%%%%%%%%%%%%%%%%%%%%%%%
%%%%%%%%%%%%%%%%%%%%%%%%%%%%%%%%%%%%%%%%%%

%Here, we have shown that the SHC in Ta exhibits such an
%anomalous $\gamma$-dependence in the low resistive regime based on the NRL-TB model.

Here, we have shown that the presence of almost degenerate anticrossing points in Ta gives rise to the anomalous $\gamma$-dependence in the low resistive regime based on the NRL-TB model.
In this exceptional situation, $\sxy^{zIIb}$ plays an significant role.
Except for this special case, however, SHC is mainly given by the Fermi surface term $\sxy^{zI}$.
We may have to confirm this anomalous behavior in Ta by checking the 
accuracy of the NRL-TB model in detail.

%%%%%%%%%%%%%%%%%%%%%%%
%Discussions
%%%%%%%%%%%%%%%%%%%%%%%
\section{\label{sec:level5} DISCUSSIONS}
%%%%%%%%%%%%%%%%%%%%%
%Effective Magnetic Flux
%%%%%%%%%%%%%%%%%%%%%
\subsection{\label{subsec:level5-1} EFFECTIVE MAGNETIC FLUX}

%rewritten

\begin{figure}[!htbp]
\includegraphics[width=0.9\linewidth]{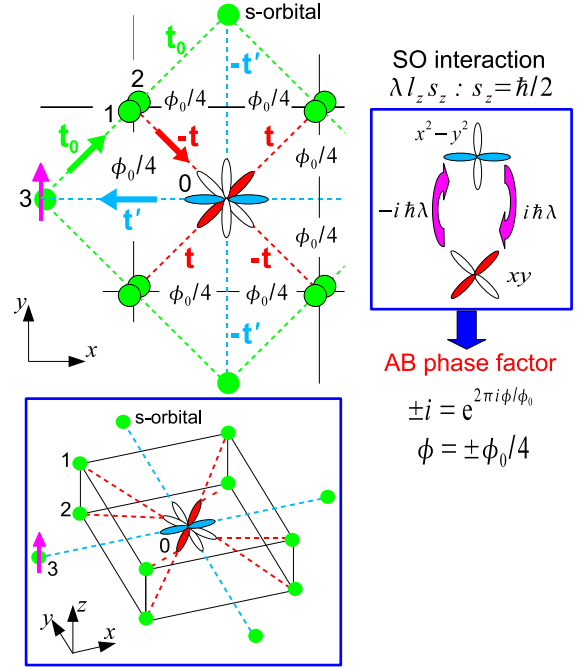}
\caption{\label{fig:AB-phase} Effective AB phase in double layer bcc model for $\uparrow$-spin electron. This is the origin of the huge Hall conductivities in various transition metals. }
\end{figure}

In the previous section, we have discussed the SHC based on multiorbital tight-binding model using the Green function method. 
In this section, we give an intuitive reason why huge SHC appears in the present multiorbital model
based on the double layer bcc model in Fig. \ref{fig:AB-phase}, which is a simplified version of the bcc structure model.
Here, we consider only $d_{xy}$-, $d_{x^2-y^2}$- and $s$-orbitals considering the fact that $d_{xy}$- and $d_{x^2-y^2}$-orbitals give the dominant contribution to the SHC in various transition metals as explained in section \ref{sec:level4}.
In Fig. \ref{fig:AB-phase}, $\pm t$ represents the hopping integrals between nearest neighbor $d_{xy}$-orbital and $s$-orbital, and $\pm t'$ is for the next nearest neighbor $d_{x^2-y^2}$-orbital and $s$-orbital. Note that both hopping integrals change their signs by rotation by $\pi/2$. $t_0$ represents the hopping integral between $s$-orbitals.

%%%%%%%%%%%%%%%%%%%%%%%%%%%%%%%%%%%%%%%%%%
%%%%%%%%%%%%%%%%%%%%%%%%%%%%%%%%%%%%%%%%%%

First, we explain that electron can transfer from $d_{x^2-y^2}$-orbital state to $d_{xy}$-orbital state and $\textit{vice versa}$ by using SOI for $\uparrow$-spin electron $\hbar \ld \hat l_z/2$, which plays a significant role in the large SHE and OHE in transition metals \cite{Kontani-Pt}.
$| x^2-y^2 \rangle$ state is transferred to $| xy \rangle$ state by rotating the wave function around $z$-axis by $\pi/4$.
Since $\hat l_z$ is a generator of the rotaion operator about $z$-axis $\hat R_z(\theta)=\e^{-i\theta l_z}$,  the relation between $\hat l_z$ and $\hat R_z(\theta)$ is given by 
\begin{\eq}
%\mp i \hat l_z &= \hat R_z(\pm \f{\pi}{2}) \qquad \text{if} \ \ \  l_z^{2} =1, \\
\mp \f{i}{2} \hat l_z &= \hat R_z(\pm \f{\pi}{4}),
\end{\eq} 
for $l_z=\pm 2$ basis.
Therefore,
%\begin{\eq}
%\hat l_z |zx \rangle =& i \hat R_z(\f{\pi}{2})|zx \rangle = i |yz \rangle, \\
%\hat l_z |yz \rangle =& -i \hat R_z(\f{\pi}{2})|yz \rangle = -i |zx \rangle,
%\end{\eq}
%for $\hat l_z^2$=1, and
\begin{\eq}
\hat l_z |x^2-y^2 \rangle =& 2i \hat R_z(\f{\pi}{4})|x^2-y^2 \rangle = 2i |xy \rangle, \\
\hat l_z |xy \rangle =& -2i \hat R_z(\f{\pi}{4})|xy \rangle = -2i |x^2-y^2 \rangle.
\end{\eq}
As a result, we obtain the off-diagonal matrix element of SOI for $\uparrow$-spin electron as follows:
\begin{\eq}
%\langle yz |\hat l_z| zx \rangle &=& -\langle zx|\hat l_z| yz\rangle = i, \\
\langle xy |\hbar \lambda \hat l_z/2| x^2-y^2 \rangle &=& -\langle x^2-y^2 |\hbar \lambda \hat l_z/2 | xy \rangle =i \hbar \lambda. \nn \label{eq:SO-xyxy}
\end{\eq}

Figure \ref{fig:AB-phase} shows the most important process (interorbital hopping process) for SHE in real space. 
%the $\uparrow$-spin electron acquires the factor $+\hbar \ld$ since the matrix element of $l_z$ is given by $\langle xy |l_z | x^2-y^2 \rangle = -\langle x^2-y^2 |l_z | xy \rangle =2i $.
By considering the signs of interorbital hopping integrals ($\pm t$ and $\pm t'$) and matrix elements of SOI in eq. (\ref{eq:SO-xyxy}), we can verify that a clockwise (anti-clockwise) movement of a $\uparrow$-spin electron along the path ``0$\rightarrow$3$\rightarrow$1 or 2$\rightarrow$0", for example, causes the factor $+i(-i)$.
%the $\uparrow$-spin electron acquires the factor $+i$ along a triangle of a half unit cell 0$\rightarrow$(1,2)$\rightarrow$3$\rightarrow$0, for example, since the matrix element of $l_z$ is given by $\langle xy |l_z | x^2-y^2 \rangle = -\langle x^2-y^2 |l_z | xy \rangle =2i $
%we can verify that a clockwise (anti-clockwise) movement of a $\uparrow$-spin electron along any triangle of half unit cell causes the factor $+i$ ($-i$). 
This factor can be interpreted as the Aharonov-Bohm phase factor $\e^{2\pi i\phi/\phi_0}$  [$\phi_0=hc/|e|$], where $\phi$ represents the effective magnetic flux
$\phi= \oint{\bf A} d{\bf r}=\pm \phi_0/4$. 
%We note that electrons acquire effective AB phase factor when they move along any triangle path using the interorbital hopping integrals in Fig. \ref{fig:AB-phase}.
Since the effective magnetic flux for $\downarrow$-spin electron is opposite in sign, electrons with different spins move to opposite direction. Therefore, the effective magnetic flux gives rise to the SHC of order $O(\lambda)$. 
This mechanism will be realized in various multiorbital transition metals \cite{Kontani-Ru,Kontani-Pt}.

\subsection{\label{subsec:level5-3} Coherent-Incoherent Crossover of Intrinsic Hall Conductivities}

In section \ref{sec:level4}, we studied the $\g$-dependence of the SHC in Ta and W numerically. 
Therein, we have verified that a typical crossover behavior of the intrinsic SHC at $\g\sim \Delta$ is realized in many transition metals unless almost degenerate anticrossing points exist slight away from the Fermi level. These crossover behavior of $\sxy$ was shown in ref. \cite{{Kontani94},{Kontani06}}.
%agree with the results of ref. \cite{Kontani06}.
Here, we discuss the crossover behavior of intrinsic Hall conductivities analytically by dividing into 3 regimes with respect to $\g$: $\g\ll\Delta$, $\Delta \ll \g \ll W$, and $W \ll \g$, where $\Delta$ and $W$ represent the band-splitting near the Fermi level and the bandwidth, respectively. The first regime corresponds to the low resistive regime, and the second to the high resistive regime. Here, we discuss the regime $W\ll\g$ only briefly since the Ioffe-Regel condition $\g\sim W$ is violated.
%Therefore, hereafter, we don't discuss the $\g$-dependence of SHC in the regime $\g\ll W$.

%%%%%%%%%%%%%%%%%%%%%%%%%%%%%%%%%%%%%%%%%%%
\begin{table}[!htbp]
\caption{\label{table:crossover} $\g$-dependence of the Fermi surface term $\sxy^{I}$, Fermi sea terms $\sxy^{IIa}, \sxy^{IIb}$, and the longitudinal conductivity $\sigma_{xx}$.}
\begin{ruledtabular}
\begin{tabular}{l|c|c||c}
  & $\g\ll \Delta$ (low $\rho$) & $\Delta\ll\g\ll W$ (high $\rho$) & $W \ll\g$ \\ \hline
$\sxy^{I}$ & $\g^0$ \ \ & $\g^{-2}$ \ \ & $\g^{-3}$  \\ 
$\sxy^{IIa}$ & $\g^0$  \ \  &  $\g^{0}$  \ \  & $\g^{-1} $ \\
$\sxy^{IIb}$ & $\g^0$  \ \  & $\g^{0}$  \ \  & $\g^{-1} $ \\ \hline
$\sigma_{xx}$ & $\g^{-1}$  \ \  & $\g^{-1}$  \ \  & $\g^{-2} $ \\
\end{tabular}
\end{ruledtabular}
\end{table}
%%%%%%%%%%%%%%%%%%%%%%%%%%%%%%%%%%%%%%%%%%%%%%
Now, we analyze eqs. (\ref{eq:sxyI-1}), (\ref{eq:IIa}) and (\ref{eq:IIb}) to obtain the $\g$-dependence of the Fermi surface term $\sxy^{I}$ and the Fermi sea term $\sxy^{II}$. 
The $\g$-dependence of  $\sxy^{I}$ is estimated by analyzing eq. (\ref{eq:sxyI-1}) as follows for $\g \ll W$, by considering the following relationship:
\begin{align}
\text{Im} &\ltk \f{1}{(E^{l}-i\g )(E^{m}+i\g )} \rtk = \f{\g (E^{l}_{\bk}-E^{m}_{\bk})}{((E^l_{\bk})^2+\g^2 ) ((E^m_{\bk})^2+\g^2 ) } \label{eq:I-Im0} \\ 
& \qquad\qquad \approx \f{\pi}{\g}\f{\g (E^m_{\bk}-E^l_{\bk}) (\delta (E^{l}_{\bk}) + \delta(E^m_{\bk} ))}{(E^m_{\bk}-E^l_{\bk})^2+\g^2}. \label{eq:I-Im}
\end{align}
After $\bk$-summation, eq. (\ref{eq:I-Im}) is proportional to $\g^0$ for $\g\ll\Delta$, and it is proportional to $\g^{-2}$ for $\Delta \ll \g \ll W$. On the other hand, in the regime $W\ll\g$, we can estimate eq. (\ref{eq:I-Im0}) as
\begin{align}
\sum_{\bk} \f{\g (E^{l}_{\bk}-E^{m}_{\bk})}{((E^l_{\bk})^2+\g^2 ) ((E^m_{\bk})^2+\g^2 ) } &\sim \f{\g}{\g^4} \sum_{\bk} (E^l_{\bk}-E^m_{\bk}) \nn 
&\sim \g^{-3}.
\end{align}
In a similar way, the $\g$-dependence of the Fermi sea terms $\sxy^{IIa}$ and $\sxy^{IIb}$ can be estimated from eqs. (\ref{eq:IIa}) and (\ref{eq:IIb}), respectively. 
The longitudinal conductivity $\sigma_{xx}$ is given by 
\begin{align}
\sigma_{xx}= \f{1}{2\pi N} \sum_{\bk} &\text{Tr} \ldk \hat J^C_x \hat G^R \hat J^C_x \hat G^A \right. \nn
&\left. -\f{1}{2} \ltk \hat J^C_x \hat G^R \hat J^C_x \hat G^R + \langle R \leftrightarrow A \rangle \rtk \rdk. 
\end{align}
In Table \ref{table:crossover}, the $\g$-dependences of $\sxy^{I}$, $\sxy^{IIa}$, $\sxy^{IIb}$, and $\sigma_{xx}$ are shown. 
%In this table, we don't discuss the $\g$-dependence of SHC in the regime $\g\ll W$ since the present calculation doesn't take the localization effect of electron into account. 
%From eqs. (\ref{eq:sxyI-1}), (\ref{eq:IIa}) and (\ref{eq:IIb}), we can estimate the $\g$-dependence of the Fermi surface $\sxy^{I}$, $\sxy^{IIa}$, $\sxy^{IIb}$, and the longitudinal conductivity $\sigma_{xx}$ as shown in Table \ref{table:crossover}. Here, the regime $\g\ll\Delta$ corresponds to the low resistive regime and $\Delta\ll\g \ll W$ corresponds to the high resistive regime, where $\Delta$ is the minimum band-splitting near the Fermi level and $W$ is the bandwidth. In the regime $W\ll\g$, the system may lose metallic behavior since the Ioffe-Legel limit $\g\sim W$ is violated, which will discussed later. Therefore, in the metallic systems, the crossover behavior of the intrinsic Hall conductivities are seen around $\g\sim\Delta$. 
In the metallic systems, the relations $\sxy\approx\sxy^{I}$ (Fermi surface term) and $|\sxy^{I}|\gg|\sxy^{II}|$ hold well for a wide range of $\g$ since the Fermi sea terms $\sxy^{IIa}$ and $\sxy^{IIb}$ almost cancel each other \cite{Kontani06}. Therefore, we discuss the $\g$-dependence of the Fermi surface term in detail. 
From Table \ref{table:crossover}, the Fermi surface term $\sxy^{I}$ is independent of $\g$ in the low resistive regime, whereas $\sxy^{I}$ decreases approximately in proportion to $\g^{-2}$ in the high resistive regime. On the other hand, the longitudinal conductivity $\sigma_{xx}$ decreases approximately in proportion to $\g^{-1}$ in both low and high resistive regime. Therefore, the coherent-incoherent crossover behavior of intrinsic Hall conductivities at $\g\sim\Delta$ are reproduced by considering the Fermi surface term $\sxy^{I}$ correctly, as reported in refs. \cite{{Kontani94},{Kontani06}}: 
\begin{align}
&\sxy\propto \sigma_{xx}^{0} \qquad \text{for}  \ \g\ll\Delta, \label{eq:sxy-l} \\ 
&\sxy\propto\sigma_{xx}^2 \qquad \text{for} \ \Delta\ll\g\ll W. \label{eq:sxy-h}
\end{align}

In the case of $W\ll \g$, the relation $\sxy\propto \sxx^{1.5} \left( \propto \g^{-3} \right)$ holds.
However, this relation is not reliable since the Ioffe-Regel condition $(W/\gamma\sim E_F \tau \leq 1)$ is violated in this regime \cite{S.Onoda}.

Finally, we comment on the $\sxy^{IIb}$ term, which is called the Berry curvature term. In electron gas models, the relation $\sxy=\sxy^{IIb}$ holds for $\g=+0$ since $\sxy^{I}+\sxy^{IIa}=0$ \cite{{Onoda},{Sundaram}}. However, $\sxy\neq\sxy^{IIb}$ for finite $\g$, and the crossover behavior cannot be explained by analyzing $\sxy^{IIb}$, which is shown in Table \ref{table:crossover}.
In conlusion, the relationship $\sxy\approx\sxy^{I}$ and $|\sxy^{I}|\gg|\sxy^{II}|$ hold well in the real metallic systems, and the correct crossover behavior given by eqs. (\ref{eq:sxy-l}) and (\ref{eq:sxy-h}) are reproduced by the Fermi surface term $\sxy^{I}$.
%Compared to the Fermi surface term, $\sxy^{IIb}$ which corresponds to the Berry curvature term gives an erroneous result for the SHC in the high resistive regime.

Now, we comment on the Ioffe-Regel limit in transition metals: The Ioffe-Regel limit $l/a \sim k_F l \sim 1$ is approximately estimated as $E_F/\g \sim 1$, where $l,a,k_F,E_F$, and $\g$ represents an elastic mean-free path, a lattice constant, a Fermi wave number, a Fermi energy, and a quasiparticle damping rate,  respectively. From the band structure for Ta and Pt in the present model, we verified that $E_F$ is $\sim 1$ in Ta and $\sim 0.5$ in Pt. Therefore, the Ioffe-Regel limit lays around $\g \sim 1$ in Ta and $\g \sim 0.5$ in Pt, respectively. Since the localization effect of electron is not taken into account, the present calculation will be inadequate for $E_{F}/\g \gg 1$.

%%%%%%%%%%%%%%%%%%%%%%%%%
%Summary
%%%%%%%%%%%%%%%%%%%%%%%%%

\section{\label{sec:level6} Summary of the Present Study}

In this paper, we studied the intrinsic SHE and OHE in various 4$d$-transition metals (Nb, Mo, Tc, Ru, Rh, Pd, and Ag) and 5$d$-transition metals (Ta, W, Re, Os, Ir, Pt and Au) based on a multiorbital tight-binding model. We have derived the general expressions for the intrinsic SHC and OHC in the presence of overlap integrals given by eq. (\ref{eq:o1}).
We found that the huge SHCs in Pt $(5d^9 6s^1)$ and Pd $(4d^{10} 5s^0)$ are positive, whereas the SHCs in Ta $(5d^{3} 6s^2)$ and W $(5d^{4}6s^2)$ take large negative values. We also found that the SHC changes smoothly with the electron number $n=n_s+n_d$, regardless of the changes of the crystal structure.
Among the 4$d$- and 5$d$-transition metals, the magnitude of SHC in Pt shows the largest value in the low resistive regime. However, the magnitude of SHC in Ta and W exceeds that in Pt in the high resistive regime. Therefore, large negative values of SHCs in Ta and W will be observed even in the high resistive samlpes.
In this paper, we also calculated the SHC for $n=7,8$ (hcp structure).
% The SHE for the hcp structure will be studied more in detail in the later publication \cite{Kontani-hcp}.
%We have recognized that the SHC changes its sign around $n$=7,8.

We also showed that the CVC due to the local impurity potential can be safely neglected in calculating the SHC and OHC in the present model.
The obtained SHCs in various transition metals are sensitive to the changes of the chemical potential $\mu$, which reflect the multiband structure around the Fermi level $\mu$. This suggests that the intrinsic SHC can be controlled by composing alloys.
As for $\g$-dependences of SHC and OHC, we obtained the coherent-incoherent crossover behaviors in many transition metals by calculating both Fermi surface term and Fermi sea term on the same footing: $\sxy^z$ is independent of $\g$ in the low resistive regime where $\g\ll \Delta$, whereas $\sxy^z$ decreases approximately in proportion to $\rho^{-2}$ in the high resistive regime. The physical meaning of the crossover behavior can be explained as follows:
In the low resistive rigime, SHC is proportional to the lifetime of the interband excitation $\hbar/\Delta$ since it is caused by the interband particle-hole excitation induced by the electric field \cite{{KL},{Murakami-SHE},{Sinova-SHE},{Kontani06},{Kontani-Ru},{Kontani94}}. However, in the high resistive regime, SHC decreases with $\g$ since the quasiparticle lifetime $\hbar/\g$ becomes shorter than $\hbar/\Delta$ \cite{{Kontani06},{Kontani94}}.

Here, we comment on the effect of Coulomb interaction on the SHC and OHC. 
In the microscopic Fermi liquid theory,
the Coulomb interaction is renormalized to the self-energy correction and the CVC.
The renormalization factor due to the self-energy,
$z=\left( 1- \left. \f{\rd \Sigma(\w)}{\rd \w} \right|_{\w=0} \right)^{-1}$
exactly cancels in the final formulas of the SHC and OHC, eqs. (\ref{eq:sxyI-1}), (\ref{eq:IIa}) and (\ref{eq:IIb}).
As shown in ref. \cite{Kontani94},
the CVC due to the Coulomb interaction does not cause the skew scattering . 
Therefore, $\gamma$-dependences of intrinsic SHC and OHC are unchanged
by the CVC due to Coulomb interactions.
However, it is well-known that the CVC causes various anomalous transport 
phenomena in the vicinity of the magnetic quantum critical points (QCP)
\cite{Kontani-Hall,Kontani-MR,Kontani-S,Kontani-Nernst,Kontani-Yamada}.
In the same way, prominent CVC near the magnetic QCP may
cause novel temperature dependence of the SHC and OHC.
This is an important future problem.
%However, if an accidental degenerate point exists slightly away from the Fermi level, an anomalous $\g$-dependence of SHC can occur: the SHC decreases as $\g$ decreases even in the low resisitive regime.

%ADD
Owing to the present study of SHE and OHE in various transition metals, it has been revealed that the huge SHE and OHE are ubiquitous in multiorbital $d$-electron systems. 
In \S \ref{subsec:level5-1}, we have discussed that
the origin of these huge SHE is the ``effective AB phase" induced by the atomic SOI with the aid of interorbital hopping integrals \cite{{Kontani-Ru},{Kontani-Pt}}.
%We have also discussed in \S \ref{subsec:level5-2} that the huge OHC
%comes from the asymmetric $s$-$d$ hybridization independently of the SOI,
%and the SHE is induced by the OHE in the presence of the SOI.
%Both explanations offer us a useful intuitive understanding of the SHE and OHE
%in transition metals.
%Large effective magnetic flux $\phi_0/2$ ($-\phi_0/2$) per unit cell exists for $\downarrow$-spin ($\uparrow$-spin) electrons, where $\phi_0=hc/|e|$. 
The present study strongly suggests that ``giant SHE and OHE " will be seen ubiquitously in multiorbital $f$-electron systems with atomic orbital degrees of freedom. 
These facts will enable us to constract efficient spintronics or orbitronics 
devices made of transition metals. 
Furthermore, in $f$-electron systems, a larger SHE may appear compared to that of $d$-electron systems since the angular momentum of atomic orbital is larger and the band splitting near the Fermi surface is smaller.

In the presence of anticrossing bands, 
huge SHC can be realized when the Fermi level lies inside the gap
induced by SOI \cite{{Murakami-qSHC},{Kane}}.
In this case, $\sigma_{xy}^I=0$ and $\sigma_{xy}^{II}$ takes a large (and almost quantized) value.
In the present study, however, we could not find any elemental metals in which the Dirac cone type band structure 
has a dominant contribution to SHE in the low resistive regime. 
The large SHCs in transition metals are mainly given by the Fermi surface term, $\sigma_{xy}^I$.
Therefore, the existence of Dirac point is not a necessary condition for large SHE: As shown in Fig. \ref{fig:Pt-disp}, the band structure where the band-splitting near the Fermi level is small is significant for large SHE.
However, only in the case of Ta, almost degenerate anticrossing points that exists slightly away from the Fermi level gives rise to an anomalous $\g$-dependence in the low resistive regime.
The anomalous $\g$-dependence of SHC in Ta may be realized.

Finally, we discuss the quantitative accuracy of the obtained numerical results, 
which depends on the accuracy of the band structure of the model near the Fermi level.
SHC and OHC depend on the multiband structure near the Fermi level
with small interband splitting $\Delta$, and they are proportional to 
$\Delta^{-1}$ according to eq. (\ref{eq:sxyI-1}).
According to ref. \cite{Papas}, the possible error in the NRL-TB is about 0.002-0.004 Ry,
which is much smaller than $\Delta$ ($\Delta \sim 0.035$Ry in Pt) that gives the minimum energy scale 
in the intrinsic Hall effect.
Therefore, it is expected that the NRL-TB model is accurate enough 
to derive qualitatively reliable results of SHC and OHC.
Thus, the overall $n$-dependence of the SHC in Fig. \ref{n-dep} will be reliable.
\begin{acknowledgments}
We are grateful to D. A. Papaconstantopoulos and I. Mazin for useful
comments and discussions on the NRL-TB model.
We also thank H. Aoki, H. Fukuyama, M. Ogata, and E. Saitoh  for fruitful discussions.
This study has been supported by Grants-in-Aid for Scientific
Research from the Ministry of Education, Culture,
Sports, Science and Technology of Japan.
Numerical calculation were performed at the facilities
of the Supercomputer Center, ISSP, University of Tokyo. 
\end{acknowledgments}

\appendix

%%%%%%%%%%%%%%%%%%%%%
%Appendix A
%%%%%%%%%%%%%%%%%%%%%
\section{\label{level:App-A} DERIVATION OF EQ. (\ref{eq:nq})}

Here, we derive eq. (\ref{eq:nq}). $n(\bq)$ is given by Fourier transform of electron number density $n(\br)$ as follows:
\begin{\eq}
n(\bq) &=& \int d\br n(\br) \e^{-i \bq \br}  \nn
         &=& \int d\br \psi^{\dagger}(\br) \psi(\br) \e^{-i \bq \br}  \nn
         &=& \sum_{\bk,\bk'} \sum_{\alpha,\beta} \ldk \int d\br \phi^{\ast}_{\bk \alpha}(\br) \phi_{\bk' \beta}(\br) \e^{-i \bq \br} \rdk c^{\dagger}_{\bk \alpha} c_{\bk' \beta}, \nonumber \\ \label{eq:nq-def}
\end{\eq}
where $\psi(\br)$ represents the electron field operator and this operator can be expanded with the atomic wave function $\phi_{\bk l}(\br)$ as follows:
\begin{\eq}
\psi(\br) = \sum_{\bk,\alpha} \phi_{\bk \alpha}(\br) c_{\bk \alpha}.
\end{\eq}
We used this relation to transfer from the second row to the third row in eq. (\ref{eq:nq-def}).

From Bloch's theorem, the atomic wave function can be rewritten as
\begin{\eq}
\phi_{\bk \alpha}(\br) = \f{1}{\sqrt{N}} u_{\bk \alpha}(\br) \e^{i\bk\cdot\br},
\end{\eq} 
where
\begin{\eq}
u_{\bk \alpha}(\br) = \sum_{i} \e^{i \bk(\bR_i - \br)} \phi_{\alpha} (\br-\bR_i).
\end{\eq}
Then, it is straight foward to show that the equation in square bracket in eq. (\ref{eq:nq-def}) is rewritten as
%Then, [ $\cdot$ ] in eq. (\ref{eq:nq-def}) can be calculated as
\begin{\eq}
%&\int d\br \phi^{\ast}_{\bk \alpha}(\br) \phi_{\bk' \beta} \e^{-i \bq \cdot\br} \nn
%&=&\f{1}{N} \int d\br u^{\ast}_{\bk \alpha}(\br) u_{\bk' \beta}(\br) e^{i(-\bk +\bk'-\bq)\cdot\br} \nn
%&=& \f{1}{N} \int_{unit} d\br u^{\ast}_{\bk \alpha}(\br) u_{\bk' \beta}(\br) \sum_{\bR_i} \e^{i (-\bk +\bk'-\bq)\cdot\bR_i} \nn
\delta(\bk+\bq-\bk') \int_{unit} d\br u^{\ast}_{\bk \alpha}(\br) u_{\bk' \beta}(\br).
\end{\eq}
%where
%\begin{\eq}
%u_{\bk \alpha}(\br) = \sum_{i} \e^{i \bk(\bR_i - \br)} \phi_{\alpha} (\br-\bR_i).
%\end{\eq}
%
Therefore, eq. (\ref{eq:nq-def}) is rewriten as
\begin{\eq}
\sum_{\bk,\alpha,\beta} \ldk \int_{unit} d\br u^{\ast}_{\bk \alpha}(\br) u_{\bk + \bq,\beta}(\br) \rdk c^{\dagger}_{\bk \alpha}c_{\bk+\bq,\beta}.  \label{eq:nq-mid}
\end{\eq}

Here, we expand [ $\cdot$ ] in eq. (\ref{eq:nq-mid}) in powers of $q/2 \ (q=|\bq|)$ as follows: 
\begin{\eq}
%&\int_{unit} d\br u^{\ast}_{\bk \alpha}(\br) u_{\bk +\bq,\beta}(\br) \nn
&&\int_{unit} d\br u^{\ast}_{\bk+\bq/2-\bq+2,\alpha}(\br) u_{\bk+\bq/2+\bq/2,\beta}(\br) \label{eq:nq-ex0} \\ 
&=&\int_{unit} d\br u^{\ast}_{\kt \alpha} u_{\kt \beta} + \f{q}{2} A_{\alpha\beta} + O\lk \f{q}{2} \rk, \label{eq:nq-ex}
\end{\eq}
where $\kt$ is given by $\kt= \bk+\bq/2$, and $\f{q}{2} A_{\alpha\beta}$ is given by
\begin{align}
&\int_{unit} d\br \sum_{i,j} \nn
&\times\ldk \f{\bq}{2}\cdot (\bR_i -\br)\e^{-i\kt (\bR_i -\br)} \phi^{\ast}_{\alpha}(\br-\bR_i)\e^{i\kt (\bR_j -\br)} \phi_{\beta}(\br-\bR_j)  \right.  \nn
  &\left. + \f{\bq}{2}\cdot (\bR_j - \br) e^{-i\kt (\bR_i -\br)}\phi^{\ast}_{\alpha}(\br-\bR_i)\e^{i\kt (\bR_j -\br)} \phi_{\beta}(\br-\bR_j)  \rdk.  \label{eq:Aab}
\end{align}

First, the first term on the right hand side in eq. (\ref{eq:nq-ex}) can be calculated as follows:  
\begin{\eq}
&& \int_{unit} d\br \sum_{i,j} \e^{-i \kt (\bR_i - \br)} \phi^{\ast}_{\alpha}(\br - \bR_i) \e^{i \kt (\bR_j -\br )} \phi_{\beta}(\br - \bR_j)  \nn
&=& O^{-1}_{\alpha\beta} (\kt).
\end{\eq}

Next, we show that $\f{q}{2}A_{\alpha\beta}$ given by eq. (\ref{eq:Aab}) vanishes
by rewritting eq. (\ref{eq:Aab}) as follows:
\begin{align}
%&&\int_{unit} d\br \sum_{i,j} \sum_{\kt'} \delta_{\kt\kt'} \nn
%&&\times\ldk \f{\bq}{2}\cdot (\bR_i -\br)\e^{-i\kt (\bR_i -\br)} \phi^{\ast}_{\alpha}(\br-\bR_i)\e^{i\kt' (\bR_j -\br)} \phi_{\beta}(\br-\bR_j)  \right.  \nn
%  &&\left. + \f{\bq}{2}\cdot (\bR_j - \br) e^{-i\kt (\bR_i -\br)}\phi^{\ast}_{\alpha}(\br-\bR_i)\e^{i\kt' (\bR_j -\br)} \phi_{\beta}(\br-\bR_j)  \rdk \nn
&\sum_{\kt'} \delta_{\kt\kt'} \f{\bq}{2}\cdot\lk \f{\rd}{\rd \kt} - \f{\rd}{\rd \kt'} \rk \nn
&\times\ldk \int_{unit} d\br \sum_{i,j} e^{-i\kt (\bR_i - \br)} \phi^{\dagger}_{\alpha} (\br -\bR_i) e^{i\kt' (\bR_j - \br)} \phi_{\beta}(\br - \bR_j) \rdk \nn
&= \f{\bq}{2N} \cdot\lk \f{\rd}{\rd \kt} O^{-1}_{\alpha\beta} -\f{\rd}{\rd \kt} O^{-1}_{\alpha\beta}  \rk=0.
\end{align}

Therefore, the final result for $n(\bq)$ is given by
\begin{\eq}
n(\bq) = \sum_{\bk,\alpha,\beta} O^{-1}_{\alpha\beta} (\bk) c^{\dagger}_{\bk-\bq/2, \alpha} c_{\bk+\bq/2,\beta}.
\end{\eq}
which is exact up to $O(q)$.

\section{\label{level:App-C} Definitions of Spin and Orbital Current Operators}

In the present study, we assume that the spin and orbital current operators are given by eqs. (\ref{eq:Js}) and (\ref{eq:Jo}) according to literatures \cite{{Inoue-SHE},{Guo}}. 
Here, show the validity for these definitions in a microscopic way. Since the spin and the orbital operators are not conserved in the present model with SOI, we can not  define spin and orbital current operators from the continuity equations. However, it is possible to make a natural definition for each current operator, as follows.

First, we consider the spin current operator. Since the SOI in the present model is local, we can virtually apply a magnetic field (vector potential) to $\uparrow$-spin and $\downarrow$-spin electrons separately. Here, we denote the vector potential for $\uparrow$-spin and $\downarrow$-spin as $\bm{A}_{\uparrow}$ and $\bm{A}_{\downarrow}$, respectively. By considering a transformation $\bk_{\sigma} \rightarrow \bk_{\sigma} -e \bm{A}_{\sigma}$, where $\sigma$ is a spin index, the $x$-component of current operators for $\uparrow$-spin and $\downarrow$-spin electrons are given by 
\begin{\eq}
J_{\uparrow x} = \f{\rd \hat H}{\rd A_{\uparrow x}} = 
-e \left(
\begin{array}{cc}
v_x & 0 \\
0 & 0
\end{array}
\right), \\
J_{\downarrow x} = \f{\rd \hat H}{\rd A_{\downarrow x}} = 
-e \left(
\begin{array}{cc}
0 & 0 \\
0 & v_x
\end{array}
\right).
\end{\eq}
Therefore, the natural definition of spin current operator in the present model is given by 
\begin{align}
J^S_{x} = \f{1}{(-e)} \left( J_{\uparrow x} - J_{\downarrow x}  \right).
\end{align} 
This expression is equivalent to eq. (\ref{eq:Js}).
We note that above discussion can not be applied to systems in the presence of non-local SOI, such as Rahsba type SOI.

Next, we consider the orbital current operator. 
%Since the electric charge conserves in the present model, the charge current operator can be derived from the equation of motion as follows:
The charge current operator $J^C_{x} =J_{\uparrow x} + J_{\downarrow x}$
is expressed in the real space representation as
\begin{align}
\bm{J}^{C} &=\sum_{im,jm'} \bm{J}^{C}_{im,jm'} ,
 \\
\bm{J}^{C}_{im,jm'} &=  -e(\bm{r}_i -\bm{r}_{j}) \cdot t_{im,jm'} (c^{\dagger}_{im}c_{jm'}-c^{\dagger}_{jm'}c_{im}),
\end{align}
where $i$ is the position of $i$th lattice point, $m$ represents the eigenvalue of $\hat l_z$, and $t_{im,jm'}$ represents the hopping integral between $|m \rangle$ state at $i$th site and $|m' \rangle$ state at $j$th site, respectively.
Then, the natural definition of the orbital current operator will be given by
\begin{\eq}
(\bm{J}^O)_{im,jm'} = \f{1}{2(-e)} (m+m') (\bm{J}^C)_{im,jm'}, \label{eq:Jo-r} 
\end{\eq}
in the basis of $l_z=2,1,\cdot\cdot\cdot,-2$.
By perfoming Fourier transforms of eq. (\ref{eq:Jo-r}), we obtain the following expression for the orbital current operator in the present model: 
\begin{\eq}
(\bm{J}^O(\bk))_{mm'} = \f{1}{2(-e)} (m+m') (\bm{J}^C(\bk))_{m,m'}.
\end{\eq}
In general basis, the above equation can be rewritten as
\begin{\eq}
\bm{J}^{O} = \ltk \bm{J}^C,l_z \rtk/2(-e).
\end{\eq}
This expression is equivalent to eq. (\ref{eq:Jo}).
In summary, we have introduced a natural definition of the spin and orbital current
operator, and shown that they are equivalent to eqs. (\ref{eq:Js}) and (\ref{eq:Jo}), respectively.

In the same manner, if we define the spin current operator 
(in the presence of the intersite SOI) as
$(\bm{J}^S)_{im\sigma,jm'\sigma'}
 = \f{1}{2(-e)} (\sigma+\sigma') (\bm{J}^C)_{im\sigma,jm'\sigma'}$,
we can derive the spin current $\bm{J}^{S} = \ltk \bm{J}^C,s_z \rtk/2(-e)$
immediately.
This is another microscopic derivation of the spin current operator
in eq. (\ref{eq:Js}).

%%%%%%%%%%%%%%%%%%%%%%%%
%references
%%%%%%%%%%%%%%%%%%%%%%%%


\begin{thebibliography}{99}
\bibitem{S.Onoda}
S. Onoda, N. Sugimoto, and N. Nagaosa, Phys. Rev. Lett. {\bf 97} (2006) 126602.

They seem to have derived the relation $\sxy\propto \sxx^{1.6}$
from the regime $E_F \tau \lesssim 1$, which may be outside the scope of
the $T$-matrix approximation.
\end{thebibliography}
\end{document}